\documentclass[12pt,a4paper]{article}
\pdfoutput=1
\usepackage{a4wide}
\usepackage[centertags]{amsmath}
\usepackage{amssymb}
\usepackage{cite}
\usepackage{ulem}
\usepackage{float}
\usepackage{amsmath}
\usepackage[pdftex]{graphicx,color} 
\usepackage{soul}
\allowdisplaybreaks
\numberwithin{equation}{subsection}
\begin{document}
	
\newenvironment{gleichung}{\begin{equation}\begin{aligned}}{\end{aligned}\end{equation}\noindent} 

\renewcommand{\thefootnote}{\fnsymbol{footnote}}

\thispagestyle{empty}

\begin{center}
  {\Large  \bf Supersymmetric Black Holes and the SJT/$\mathbf{nSCFT_1}$
    Correspondence}  
\end{center}

\vspace*{1cm}

\centerline{Stefan F{\"o}rste$^1$, Andreas Gerhardus$^2$\footnote{AG contributed to this work while he was still at the Bethe Center for Theoretical Physics and the Physikalisches Institut der Universit\"at Bonn.} and Joshua Kames-King$^1$}
\vspace{1cm}

	\begin{center}{
	\it
	$^1$Bethe Center for Theoretical Physics\\
	{\footnotesize and}\\
	Physikalisches Institut der Universit\"at Bonn,\\
	Nussallee 12, 53115 Bonn, Germany
	}
	\end{center}
	\begin{center}{
	\it
	$^2$German Aerospace Center\\
	Institute of Data Science\\
	07745 Jena, Germany
	}
	\end{center}

\vspace*{1cm}

\centerline{\bf Abstract}
\vskip .3cm

We consider 1/4 BPS black hole solutions of ${\cal N}=2$ gauged supergravity in $AdS_4$. The near horizon geometry is $AdS_2 \times S^2$ and supersymmetry is enhanced. In the first part of the paper we choose a moment map, which allows the embedding of this supergravity solution into a sugra theory with a hypermultiplet. We then perform the s-wave reduction of this theory at the horizon and determine the dilaton multiplet, which couples to both metric and gravitino fluctuations. In the second part we work with Euclidean axial $\mathcal{N}=(2,2)$ JT supergravity and show how to add gauged matter in form of  covariantly twisted chiral and anti-chiral multiplets. We demonstrate how to reduce the on-shell action to boundary superspace. We compare both theories and calculate the fourpoint function by integrating out gravitons, gravitini and photons for the s-wave setting and by use of the Super-Schwarzian modes in the JT theory.
\vskip .3cm

\newpage

\renewcommand{\thefootnote}{\arabic{footnote}}
\setcounter{footnote}{0} 

\section{Introduction}

In the present paper we will carry out a supersymmetric extension of studies presented in \cite{Nayak:2018qej,Moitra:2018jqs}. Extremal black holes contain an $AdS_2$ factor in the near horizon limit in which finite energy excitations decouple \cite{Maldacena:1998uz}. 
In order to capture also those it has been proposed in \cite{Almheiri:2014cka} to add Jackiw-Teitelboim gravity \cite{Jackiw:1984je,Teitelboim:1983ux} to the bulk action. By integrating out bulk fields this can be related to an effective one dimensional theory whose Lagrangian is given by the Schwarzian derivative of the boundary curve \cite{Jensen:2016pah,Maldacena:2016upp,Engelsoy:2016xyb,Cvetic:2016eiv}. This is also the effective Lagrangian arising in the strong coupling limit of the SYK model   \cite{Sachdev:1992fk,Kitaev,Maldacena:2016hyu}. (For reviews on the SYK/JT correspondence see e.g.\ \cite{Sarosi:2017ykf,Rosenhaus:2018dtp}.) In the present paper we will be interested in supersymmetric extensions. On the SYK side ${\cal N} \in \{ 1, 2\}$ extensions have been presented in \cite{Fu:2016vas}. The effective Lagrangian at strong coupling is given by the corresponding super-Schwarzian derivatives.
The ${\cal N} = (1,1)$ extension of JT gravity on manifolds without a boundary is given in \cite{Chamseddine:1991fg}. The inclusion of a boundary term, the extension to ${\cal N} =(2,2)$ and the relation to super-Schwarzians is presented in \cite{Astorino:2002bj,Forste:2017kwy,Forste:2017apw}. In the present paper, we will be interested in the ${\cal N} = \left( 2,2\right)$ configuration of which further aspects have been studied in e.g.\ in 
\cite{Stanford:2017thb,Mertens:2017mtv,Kanazawa:2017dpd,Murugan:2017eto,Yoon:2017gut,Peng:2017spg,neues,Bulycheva:2018qcp,Peng:2018zap,Chang:2018sve,Cardenas:2018krd,Berkooz:2020xne,Peng:2020euz}.

In \cite{Nayak:2018qej,Moitra:2018jqs} (see also e.g.\ \cite{Castro:2018ffi,Moitra:2019bub}) the relation of the $nAdS_2/nCFT_1$ correspondence to higher dimensional black holes is investigated in more detail. In \cite{Nayak:2018qej} an extremal and near extremal Reissner Nordstr{\"o}m $AdS_4$ black hole are considered. 
The authors compute the four point function of conformal primaries in a dual $CFT_3$ in different ways. Following \cite{Liu:1998ty} by adding a probe massive free scalar (dual to the primary under consideration) and integrating out the induced metric perturbations. This yields an expression quartic in the scalars (quadratic in energy momentum tensor components) i.e.\ quartic in the sources for the primaries in the dual CFT$_3$. Dimensional reduction (for spherically symmetric configurations and small frequences) relates this calculation to a calculation performed in the $nAdS_2/nCFT_1$ scheme. Results obtained by integrating out Schwarzian modes match. 

In the present paper we will consider a 1/4 BPS solution of gauged ${\cal N} =2$ 4d supergravity \cite{Klemm,Hristov:2010ri}. This solution represents a magnetically charged black hole with $AdS_4$ asymptotics. In the near horizon limit supersymmetry is enhanced corresponding to ${\cal N} = (2,2)$ in two dimensions. The probe should now not only preserve spherical symmetry but also supersymmetry. This can be achieved by adding a hypermultiplet along the lines of \cite{Hristov:2010eu}. 

The paper is organised as follows. Section two is devoted to the four dimensional picture and its near horizon reduction. In section \ref{sec:nohyp} we review the sugra solution \cite{Klemm,Hristov:2010ri} which does not contain hypermultiplets. 
This solution represents a black hole with $AdS_4$ asymptotics and an $AdS_2 \times S^2$ near horizon geometry. 
General techniques for adding a hypermultiplet \cite{Hristov:2010eu} are applied in section \ref{sec:withhyp}.
Section \ref{sec:hyperfluc} discusses the dimensional reduction in the near horizon limit, in $s$-wave approximation. In section \ref{sec:4pt4d} we compute four point functions in a dual CFT following \cite{Liu:1998ty,Nayak:2018qej}. That is, we integrate out metric fluctuations, gravitini fluctuations and gauge field fluctuations in a limit in which first corrections to the $S^2$ radius have been added to the near horizon limit. For the gravitini we have to impose further projections such that super currents are conserved in that limit. 

Section three is devoted to the $nAdS_2$/$nSCFT_1$ perspective on the considerations of section two. 
A natural choice for the two dimensional theory would be what we obtained from dimensional reduction in the near horizon region of the four dimensional black hole solution. However, we will twist this slightly. Firstly, we switch to Euclidean signature corresponding to the choice in \cite{Forste:2017apw}.
Another twist is performed for the following reason. We want to associate the integrating out of bulk modes (such as the graviton) to integrating out super-Schwarzian modes in the effective one dimensional dual. One of the super-Schwarzian modes corresponds to the two dimensional graviphoton. This is the gauge field of a Kaluza-Klein $U(1)$ when reducing from four to two dimensions. (The four dimensional ${\cal N}=1$ gravity multiplet does not contain a graviphoton.) The fluctuating hypermultiplet in the four dimensional setup is charged under a combination of the ${\cal N}=2$ graviphoton and an extra $U(1)$. In two dimensions this will correspond to an extra vector multiplet. We twist the charge of the probe matter in two dimensions such that it is charged under the graviphoton instead of an extra $U(1)$. In sections \ref{subsection:minimalsugra} and \ref{sec:JTsugra}
we review the ${\cal N} =\left( 2,2\right)$ extensions of JT gravity
\cite{Forste:2017apw}. After that, in section \ref{Mattercoupled2dsugra} we add a covariantly twisted chiral and anti-chiral multiplet describing probe matter. These have the same amount of degrees of freedom and the same mass as what we obtained from dimensional reduction of half the four dimensional hypermultiplet. But the covariantly twisted multiplets are charged under the two dimensional graviphoton. 
Conserved currents (energy momentum tensor, supercurrent, gauge current) share the same conservation laws with the dimensionally reduced ones and are associated to each other. Some further aspects of the relation between dimensionally reduced four dimensional theory and the considered two dimensional theory
are mentioned in section \ref{sec:4d2sdis}.
In section \ref{sec:schwMatt} the one dimensional holographic dual is considered. Supergravity is replaced by super reparametrisations with a super-Schwarzian action. Matter is coupled in a supersymmetric generalisation of the way it is presented in \cite{Maldacena:2016upp}. That is, we write down a term which generates the ${\cal N}=2$ superconformal two point functions of operators being dual to the bulk matter, in the zero temperature case. By applying a general super reparametrisation on that expression one generates the couplings to the super-Schwarzian modes. 
By integrating out (linearised) super-Schwarzian modes we obtain the expression generating four point functions of the dual superconformal operator, in section \ref{sec:schwmodes}. We express these generating functionals as two dimensional integrals containing the conserved 2d currents. Then they can be matched  with the findings of section \ref{sec:4pt4d}. We obtain agreement if we impose the same additional projection condition on the supercurrent as in \ref{sec:4pt4d}. 

In section \ref{ref:dis}, we summarize the results and discuss possible future directions. In an appendix \ref{ap:conv} we list some of the used conventions. 

\section{A supersymmetric black hole in 4d}

\subsection{Solution without hypermultiplets\label{sec:nohyp}}

In this subsection we recapitulate the 1/4 BPS magnetically charged black hole solution \cite{Klemm,Hristov:2010ri} of ${\cal N}=2$ gauged supergravity (for a review see\cite{Andrianopoli:1996cm,Freedman:2012zz,Ortin:2015hya,Hristov:2012bk}). 
Our conventions follow \cite{n=2sugra,Hristov:2012bk} and are summarized in appendix \ref{ap:conv}.
Pure gauged supergravity allows only for $AdS_4$ `black holes' with a naked singularity \cite{Romans:1991nq}.

We will first consider a solution to a theory containing the supergravity and a vector multiplet. The supergravity multiplet accomodates the vielbein $e_\mu ^a$, two gravitini $\psi_{\mu}^A$, $A\in \{ 1,2\}$, and a graviphoton $A^0 _\mu$. The vector multiplet consists of a vector $A_\mu ^1$ two gauginos $\lambda^A$ and a complex scalar $z$.  
The bosonic part of the supersymmetric Lagrangian is given by
\begin{equation}
    {\cal L} = \frac{1}{2}R\left( e\right) + g_{z z^*}\partial^\mu z\partial_\mu z^{*} +I_{\Lambda\Sigma}F^{\Lambda}_{\mu\nu} F^{\Sigma\, \mu\nu}
   +\frac{1}{2}R_{\Lambda\Sigma}\epsilon^{\mu\nu\rho\sigma}F_{\mu\nu}^\Lambda F_{\rho\sigma}^\Sigma - g^2 V\left( z, z^*\right) . 
\end{equation}
Here, $R(e)$ is the scalar curvature and $g_{zz^*}$ is the metric on a special K{\"a}hler manifold on which the scalar of the vector multiplet takes values. On a special K{\"a}hler manifold there are holomorphic sections $\left( X^\Lambda\left( z\right) , F_\Lambda\left(z\right)\right)$ where in our case $\Lambda \in \left\{ 0,1\right\}$. Further, $F_\Lambda = \partial F /\partial X^\Lambda$ where for the explicit solution in \cite{Klemm,Hristov:2010ri} the prepotential
\begin{equation}
    F = - 2 \text{i}\sqrt{ X^0 \left( X^1\right)^3}
\end{equation}
is chosen. $R_{\Lambda\Sigma}$ and $I_{\Lambda\Sigma}$ denote real respectively imaginary part of the period matrix 
\begin{equation}
    {\cal N}_{\Lambda\Sigma} = \left(\frac{\partial F_\Sigma}{\partial X^\Lambda}\right)^* + 2\text{i}\frac{\text{Im}\left( F_{\Lambda\Gamma}\right) X^\Gamma \text{Im}\left( F_{\Sigma\Delta}\right)X^\Delta}{X^E \text{Im}\left( F_{EZ}\right) X^Z} .
\end{equation}
with $F_{\Lambda\Sigma} = \partial F_\Sigma/\partial X^\Lambda$. 
The K{\"a}hler metric is expressed in terms of the K{\"a}hler potential
\begin{equation}
    g_{zz^*} = \partial_z\partial_{z^*} {\cal K} \,\,\, ,\,\,\, \text{with}\,\,\, {\cal K} = -\log \left[ \text{i}\left( \left(X^\Lambda\right)^* F_\Lambda - X^\Lambda \left( F_\Lambda\right)^*\right)\right] .
\end{equation}
The scalar potential, finally, is given by
\begin{align}
&    V =\left( g^{zz^*} f_z^\Lambda \left( f_z ^\Sigma\right)^* -3 \left( L^\Lambda\right)^* L^\Sigma\right)\xi_\Lambda \xi_\Sigma \,\,\, ,\nonumber \\&\text{with }\, f_z^\Lambda = \text{e}^{\frac{\cal K}{2}}
    \left( \partial_z + \left(\partial_z {\cal K}\right)    \right) X^\Lambda \,\,\, ,\,\,\, L^\Lambda = \text{e}^{\frac{\cal K}{2}}X^\Lambda , 
\end{align}
and the real constants $\xi^\Lambda$ are called Fayet-Iliopoulos (FI) parameters (characterising under which $U(1)$ the gravitini are charged).   
The explicit solution we are going to consider is a magnetically charged black hole.
The metric is given by
\begin{equation}
    ds^2 = U^2\left(r\right) dt^2 - U^{-2}\left( r\right) dr^2 - b^2\left( r\right)
\left( d\theta^2 + \sin^2 \theta d\varphi^2\right) ,
\label{eq:4bh}
\end{equation}
where $U$ and $b$ will be specified shortly. The non vanishing  vierbein and spin connection components are,
\begin{align}
e_\mu ^a =& \text{diag}\left(  U\left( r\right), 1/U\left(r\right), b\left( r\right), b\left( r\right) \sin\theta\right) \, ,\,\,\, 
\omega_t ^{01} =  U\partial_r U\, ,\,\,\, \omega_{\theta}^{12} = -U \partial_r b \, ,\nonumber \\ & \omega_{\varphi}^{13} = -\left(U \partial_r b \right) \sin\theta \, ,\,\,\, 
\omega_{\varphi}^{23}= -\cos\theta .
\label{eq:spincon}
\end{align}
The gauge fields have only non vanishing $\varphi$ components
\begin{equation}
    A^\Lambda_\varphi = - p^\Lambda \cos \theta\,,
\label{eq:vector}
\end{equation}
and hence the field strengths are
\begin{equation}\label{backgroundfieldstrengths}
 F_{t r}^{\Lambda}=0\,,\;\;\;\;\;F_{\theta \varphi}^{\Lambda}=\frac{p^{\Lambda}}{2}\sin{\theta}\,.
\end{equation}
Mostly we will work with the self-dual and anti-self dual field strengths defined as
\begin{equation}\begin{aligned}
 F_{\mu \nu}^{\pm\Lambda}=\frac{1}{2}(F_{\mu \nu}\mp \frac{\text{i}}{2}\epsilon_{\mu \nu \rho \sigma}F^{\rho \sigma})\,.
\end{aligned}\end{equation}
\noindent The 1/4 BPS solution reported in \cite{Klemm,Hristov:2010ri} has two Killing spinors
\begin{equation}
\epsilon_1 = \sqrt{\left( U\left( r\right)\right)}\text{e}^{-\frac{1}{4}\left( \partial_z {\cal K} \partial_r z - \partial_{z^*}{\cal K} \partial_r z^*\right)}  \epsilon_1 ^0\,\,\, ,\,\,\, \epsilon_2 = \sqrt{\left( U\left( r\right)\right)}\text{e}^{-\frac{1}{4}\left( \partial_z {\cal K} \partial_r z - \partial_{z^*}{\cal K} \partial_r z^*\right)}  \epsilon_2 ^0 ,
\end{equation}
where the $\epsilon_A ^0$ are chiral constant spinors satisfying the projection condition 
\begin{equation}\label{eq:kilspin}
    \epsilon_A = -{{\sigma^3}_A}^B \gamma_{01} \epsilon_B \,\,\, ,\,\,\, \epsilon_A = \epsilon_{AB}\gamma_0 \epsilon^B .
\end{equation}
The scalar in the vector multiplet is given by $z = X^1/X^0$ with
\begin{equation}
X^0 = \pm \frac{1}{4\xi_0} - \frac{\xi_1 \beta^1}{r\xi_0}\,\,\, ,\,\,\, X^1 = \pm \frac{3}{4\xi_1} + \frac{\beta^1}{r}
\end{equation}
with correlated signs. Asymptotically, at $r\to \infty$, $z\to \frac{ 3\xi_0}{\xi_1}$ becomes constant and so does the potential
$$ V \to \Lambda_4 = -\frac{2g^2}{\sqrt{3}}\sqrt{\xi_0{\xi_1}^3}, $$
corresponding to the radius of asymptotic $AdS_4$ geometry. The metric components in (\ref{eq:4bh}) are given by
\begin{equation}
U^2 = \text{e}^{\cal K} \left( g r + \frac{1}{2gr} -\frac{16 g}{3 r}\left( \xi_1 \beta^1\right)^2\right)^2 ,\,\,\, b^2 = \text{e}^{-{\cal K}} r^2 ,
\end{equation}
with
\begin{equation}
\text{e}^{\cal K} = \frac{1}{8\sqrt{\left(X^1\right)^3 X^0}} = \frac{2\sqrt{\xi_0\xi_1^3} r^2}{\sqrt{\left( r \mp 4 \xi_1\beta^1\right)\left( 3r \pm 4\xi_1\beta^1\right)^3}} .
\end{equation}
This is a geometry of a charged black hole, for which charges (see (\ref{eq:vector})) and mass, $M$, are all fixed in terms of the integration constant $\beta_1$, explicitly
\begin{align}
p^0 &= \frac{\mp 1}{g\xi_0} \left(\frac{1}{8} +\frac{8\left( g\xi_1\beta^1\right)^2}{3}\right)  ,\,\,\, p^1 =  \frac{\mp 1}{g\xi_1} \left(\frac{3}{8} -\frac{8\left( g\xi_1\beta^1\right)^2}{3}\right),\nonumber\\ M & = -\frac{128}{81}\Lambda_4 \left( \xi_1 \beta^1\right)^3 .
\end{align}
The metric component $U^2$ has a double zero at $r=r_h$. The position of the horizon is
\begin{equation}
r_h = \sqrt{ \frac{16}{3}\left( \xi_1\beta^1\right)^2 -\frac{1}{2g^2}} .
\end{equation}
The near horizon geometry is $AdS_2 \times S^2$ where the negative $AdS_2$ curvature overcompensates the $S^2$ curvature resulting in a negative net curvature. Supersymmetry is enhanced to 1/2 BPS corresponding to ${\cal N}= (2,2)$ in two dimensions \cite{deWit:2011gk}.

\subsection{Solution with a universal hypermultiplet\label{sec:withhyp}}

In the spirit of \cite{Nayak:2018qej} we want to switch on perturbations around a sugra solution and study the backreaction on the super geometry. We would like to do this in a supersymmetric way and to consider only spin zero and 1/2 fluctuations. We also want the perturbation to be charged under a $U(1)$ gauge symmetry such that there is a corresponding backreaction. All this can be achieved by modifying the solution of the previous section to fit into a theory with a sugra, a vector and a hypermultiplet. How to perform such a modification in general has been worked out in \cite{Hristov:2010eu}. The hypermultiplet consists of four real scalars $q^a$, $a\in \left\{ 1,2,3,4\right\}$, and two chiral fermions $\zeta_\alpha$, $\alpha \in \{ 1,2\}$. The four real scalars will be called
$$ \left( q^1, q^2, q^3, q^4\right)  = \left( R, u,v, D\right) .$$
These take values on a quaternionic-K{\"a}hler manifold which in our example is chosen to be $SU(2,1)/U(2)$ with metric
\begin{equation}
ds^2 = h_{ab}q^a q^b =\frac{1}{R^2}\left( dR^2 + R\left( du^2 + dv^2\right) + \left( dD + \frac{1}{2} u dv -\frac{1}{2}vdu\right)^2\right) .
\label{eq:quatmet}
\end{equation}
The metric $h_{ab}$ can be expressed in terms of vielbeins
(for details and conventions see \cite{Andrianopoli:1996cm, Hristov:2012bk})
\begin{equation}
h_{ab} = {\cal U}_a ^{A\alpha}{\cal U}_b ^{B\beta} \epsilon_{\beta\alpha}\epsilon_{AB} .
\end{equation}
Indices $A$ and $\alpha$ are raised and lowered with the two dimensional epsilon tensor or its transposed when they label bosonic quantities. For fermions these indices are raised and lowered by complex conjugation. A reality constraint on the vielbeins can be viewed as applying both rules simultaneously 
\begin{equation}
{\cal U}_{a, A\alpha} = \left( {\cal U}_a^{A\alpha}\right)^* = \epsilon_{AB} {\cal U}_a^{B\beta}\epsilon_{\beta\alpha} .
\end{equation}
For the metric (\ref{eq:quatmet}) the non vanishing vielbein components are 
\begin{align}
{\cal U}^{12}_R &= {\cal U}^{21}_R = \frac{1}{\sqrt{2}R},\,\,\, {\cal U}^{12}_D = - {\cal U}_D ^{21} = \frac{\text{i}}{\sqrt{2}R},\nonumber\\
{\cal U}^{21}_u &= - {\cal U}^{12}_u = \frac{\text{i}v}{2\sqrt{2} R},\,\,\, {\cal U}_u^{11} = -{\cal U}_u ^{22} = \frac{1}{\sqrt{2R}}, \label{eq:quatviel}\\
{\cal U}^{12}_v &= -{\cal U}_v^{21} = \frac{\text{i}u}{2\sqrt{2}R},\,\,\, {\cal U}_v^{11} = {\cal U}_v ^{22} = \frac{\text{i}}{\sqrt{2 R}}.\nonumber
\end{align}
The solution of \cite{Klemm,Hristov:2010ri} is invariant under the susy variations of the gravitino and the gaugino
\begin{equation}\begin{aligned}\label{gravitinogauginovariation}
 \delta_{\epsilon}\psi_{\mu A}&=\nabla_{\mu}\epsilon_{A}+2 \text{i} F_{\mu \nu}^{\Lambda -}I_{\Lambda \Sigma}L^{\Sigma}+\text{i}g S_{A B}\gamma_{\mu}\epsilon^{B}\,,\\
 \delta_{\epsilon}\lambda^{i A}&=\text{i}\partial_{\mu}z^{i}\gamma^{\mu}\epsilon^{A}+G_{\mu \nu}^{- i}\gamma^{\mu \nu}\epsilon^{A B}\epsilon_{B}+\text{i} g g^{i \bar{j}}\bar{f}^{\Lambda}_{\bar{j}}a_{\Lambda}\sigma^{A B}_{3}\epsilon_{B}\,,
\end{aligned}\end{equation}
with 
\begin{equation}\begin{aligned}\label{killingspinorcovariantderivative}
 \nabla_{\mu}\epsilon_{A}&=(\partial_{\mu}-\frac{1}{4}\omega_{\mu}^{a b}\gamma_{a b})\epsilon_{A}+\frac{\text{i}}{2}g a_{\Lambda}A_{\mu}^{\Lambda}\sigma^{3 B}_{A}\epsilon_{B}\,,\\
 G_{\Lambda \mu \nu}&= \text{Re}(\mathcal{N}_{\Lambda \Sigma})F_{\mu \nu}^{\Sigma}-\frac{1}{2}\text{Im}(\mathcal{N}_{\Lambda \Sigma})\epsilon_{\mu \nu \gamma \delta}F^{\Sigma \gamma \delta}\,,
\end{aligned}\end{equation}
provided the following BPS conditions are satisfied,
\begin{equation}\begin{aligned}\label{BPS}
 U'=&-\frac{2 L^{\Lambda}I_{\Lambda \Sigma}p^{\Sigma}}{b^2}\pm g a_{\Lambda}L^{\Lambda}\,,\\
 \frac{U}{b}b'=&\frac{2 L^{\Lambda}I_{\Lambda \Sigma}p^{\Sigma}}{b^2}\pm g a_{\Lambda}L^{\Lambda}\,,\\
 g a_{\Lambda}p^{\Lambda}=&\mp 1\,.
\end{aligned}\end{equation}
In addition to this we now demand invariance of the hyperino variation
\begin{equation}\begin{aligned}\label{hyperinovariation}
 \delta_{\epsilon}\zeta_{\alpha}=i \mathcal{U}^{\beta B}_{a}\nabla_{\mu}q^{a}\gamma^{\mu}\epsilon^{A}\epsilon_{A B} \mathcal{C}_{\alpha \beta}+2 g \mathcal{U}_{\alpha a}^{A}k_{\Lambda}^{a}\bar{L}^{\Lambda}\epsilon_{A}\,,
\end{aligned}\end{equation}
with 
\begin{equation}\begin{aligned}
  \nabla_{\mu}q&=\left(\partial_{\mu}q+g A_{\mu}^{\Lambda}k_{\Lambda}^{q}\right)\,.
\end{aligned}\end{equation}
Following the logic of \cite{Hristov:2010eu} we keep $g_{\mu \nu}, F_{\mu \nu}^{\Lambda}, z$ the same as for \cite{Klemm,Hristov:2010ri} such that \eqref{gravitinogauginovariation} is still solved. 
Parameters are fixed by the requirement
that \eqref{hyperinovariation} is also solved,
implying the conditions \cite{Klemm,Hristov:2010ri,Hristov:2010eu} 
\begin{equation}\begin{aligned}\label{hyperinokillingvectorcondition}
 k_{\Lambda}^{a}F_{\mu \nu}^{\Lambda}=0\,,\;\;\;\;P^{x}_{\Lambda}f_{i}^{\Lambda}=0\,,\;\;\;\;\epsilon^{x y z}P_{\Lambda}^{y}P_{\Sigma}^{y}L^{\Lambda}\bar{L}^{\Sigma}\,=0\,,\;\;\;\;k_{\Lambda}^{a}L^{\Lambda}=0.
\end{aligned}\end{equation}
Here, $k_\Lambda$ are Killing vectors tangent to the quaternionic-K{\"a}hler manifold resembling charge vectors of the gauged isometry. 
We consider the following Killing vectors 
\begin{equation}\label{Killingvector}
k_{\Lambda}=a_{\Lambda}\left(-v \partial_{u}+u \partial_{v}\right)\,,
\end{equation}
where $a_{\Lambda}$ functions as the FI parameter, such that the black hole background is unchanged. These Killing vectors correspond to  moment maps
\begin{equation}\label{momentmap}
 P_{\Lambda}^{x}=a_{\Lambda}\left(\frac{v}{\sqrt{R}},\frac{u}{\sqrt{R}},1-\frac{u^2+v^2}{4 R}\right)\,.
\end{equation}
Plugging \eqref{Killingvector}, \eqref{momentmap} back into \eqref{hyperinokillingvectorcondition} and \eqref{gravitinogauginovariation} we see that the following vevs are required for the hyperscalar
\begin{equation}\label{eq:hyperbg}
\langle u\rangle=\langle v\rangle =0\,,\;\;\;\;\langle R\rangle =\text{const.}\not= 0\,,\;\;\;\langle D\rangle =\text{const.}\,.
\end{equation}
Later we will consider fluctuations of $u$ and $v$. 
In a dimensionally reduced system these will be scalars of a matter multiplet which is charged under a $U(1)$ with gauge fields $A^\Lambda \sim a^\Lambda$. 
We will freeze $R$ and $D$ to their background values (\ref{eq:hyperbg}).\\
The BPS equations are not affected. In relating field strengths and mass matrices to geometrical quantities we will often need the two linear combinations of the first two BPS conditions (\ref{BPS})
\begin{equation}\label{BPSlinearcombi}
   \frac{U}{b}b'-U'=\frac{4 L^{\Lambda}I_{\Lambda \Sigma}p^{\Sigma}}{b^2}\,,\,\,\,
   U'+\frac{U}{b}b'=\pm 2 g a_{\Lambda}L^{\Lambda}\,.
\end{equation}
\subsection{Dimensional Reduction with Hypermultiplet as Source \label{sec:hyperfluc}}
We now perform the explicit s-wave reduction on the two-sphere in the near horizon geometry $AdS_2 \times S^2$. This will be carried out for the gravity multiplet, the hypermultiplet and the linearized supergravity theory coupled to the hypermultiplet as a source. The former should reproduce the JT supergravity theory in 2d and the latter should correspond to the linearized couplings in 2d. Performing this dimensional reduction will also reveal which four-dimensional fields constitute the dilaton multiplet.\\
While for a bosonic field the s-wave reduction may be implemented by assuming no dependence on spherical coordinates, for fermions an additional projection must be applied such that the degrees of freedom are reduced by half. The correct projection can for example be deduced by demanding that the vevs chosen for the hyperscalars $R,D$ are kept intact by supersymmetry variations, such that no dynamical $R,D$ fields are generated and hence $u,v$ and half of the degrees of freedom of the hyperinos constitute a proper two-dimensional multiplet.\\
By use of 
\begin{equation}
\delta q^{a}=\mathcal{U}_{\alpha A}^{a}(\bar{\zeta}^{\alpha}\epsilon^{A}+\mathcal{C}^{\alpha \beta}\epsilon^{A B} \bar{\zeta}_{\beta}\epsilon_{B})\,,
\end{equation}
the equations
$\delta R =0$ and 
    $\delta D =0$
are only fulfilled if $\zeta^{1}$ fulfills the same projection as $\epsilon^{2}$ and $\zeta^{2}$ the same as $\epsilon^{1}$\, (see the first condition in (\ref{eq:kilspin})). 
The projected spinors have only one independent component which motivates the replacement,
\begin{equation}\begin{aligned}\label{projections}
 \zeta_1 \rightarrow
 \begin{pmatrix}
  1  \\
  \text{i} \\
  \text{i}\\
  1
 \end{pmatrix}
 \zeta_1
 \,,
 \;\;\;\,\;\;\;\,
  \zeta_2 \rightarrow
 \begin{pmatrix}
  1  \\
  \text{i} \\
  -\text{i}\\
  -1
 \end{pmatrix}
 \zeta_2
 \,,
\end{aligned}\end{equation}
where in abuse of notation we have given the Grassmann variables on the right hand side of \eqref{projections} the same name as the four dimensional spinors on the left hand side.\\
The metric on $AdS_2 \times S^2$ is given by
\begin{equation}\begin{aligned}
 ds^2 = \frac{r^2}{v_1^2} dt^2 - \frac{v_1^2}{r^2} dr^2 - v_2^2\left( d\theta^2 + \sin^2 \theta d\varphi^2\right) ,
\end{aligned}\end{equation}
with the vielbein
\begin{equation}\begin{aligned}
 e^{a}_{\mu}=\text{diag}\Big(\frac{r}{v_1},\frac{v_1}{r},v_2,v_2 \sin{\theta}\Big)\,.
\end{aligned}\end{equation}
The non vanishing components of the spin connection are
\begin{equation}\begin{aligned}
 \omega_{t}^{0 1}=\frac{r}{v_1^2}\,, \;\;\;\;\;\;\;\omega_{\varphi}^{2 3}=-\cos{\theta}\,. 
\end{aligned}\end{equation}
We also note that for product space geometries, the equations of motion and the dimensionally reduced action are equivalent for the case of Einstein gravity and Maxwell theory \cite{Productspace} (barring terms which have been neglected in the approximation with at most linear contributions of the dilaton multiplet 
to the action). For the gravitini this can be easily seen to also hold: the equations of motion derived from \eqref{Gravitinidimreduced},\eqref{linearizedgravitinicurrentcoupling} match the dimensionally reduced equations \eqref{eq:j1t}--\eqref{eq:j2T} in the near horizon limit up to contributions $J^\theta _A$ containing $\psi_{\theta 
,A}$
\subsubsection{Gravity Sector}
As a first step we reduce the kinetic terms for the gravitional multiplet, which encompasses the Einstein-Hilbert term, the linear combination of the field strengths, and the gravitinos. \\
The relevant parts of the action are \cite{n=2sugra}

\begin{align}\label{4dkineticterms}
& S \supset \int \mathrm{d}^4 x \sqrt{-g}\left(-\frac{1}{2}R+\text{i}\left(\bar{\mathcal{N}}_{\Lambda \Sigma}F_{\mu \nu}^{-\Lambda}  F^{-\Lambda \mu \nu} -\mathcal{N}_{\Lambda \Sigma}F_{\mu \nu}^{+\Lambda}  F^{+\Lambda \mu \nu}\right)\right.\nonumber\\+&\left. \epsilon^{\mu \nu \lambda \sigma}\bar{\psi}_{\mu}^{A} \gamma_{\sigma}\nabla_{\nu}\psi_{\lambda A}+\text{h.c}.+F^{- \Lambda \mu \nu}\mathcal{I}_{\Lambda \Sigma}4 L^{\Sigma} \bar{\psi}_{\mu}^{A}\psi_{\nu}^{B}\epsilon_{A B}+g S_{A B}\bar{\psi}_{\mu}^{A}\gamma^{\mu \nu}\psi_{\nu}^{B}\rule{0pt}{3ex}\right)\,,
\end{align}
with the gravitino mass matrix defined as
\begin{equation}
 S_{A B}=\frac{\text{i}}{2}(\sigma_{x})_A^{C}\epsilon_{B C}P_{\Lambda}^{x}L^{\Lambda}\\
 =\frac{\text{i}}{2}(\sigma_{x_3})_A^{C}\epsilon_{B C}a_{\Lambda}L^{\Lambda}\,,
\end{equation}
where our choice of moment map (\ref{momentmap}) was applied in the second step. The covariant derivative of the gravitino is given in \eqref{killingspinorcovariantderivative}.
\noindent Furthermore, we also have to include the potential for the complex scalar in the vector multiplet linked to the FI gauging. In the full black hole solution it acts as the cosmological constant of $AdS_4$. It is given by
\begin{equation}\label{gaugescalarFI}
  V \supset  -g^2 3 L^{\Lambda} \bar{L}^{\Sigma}a_{\Lambda}a_{\Sigma}\,,
\end{equation}
where  the momentum map has already been expressed via the FI constants.\\ 
\noindent The dimensional reduction of the Einstein-Hilbert term to two dimensions was performed in \cite{Nayak:2018qej}. Assuming a static, spherically symmetric metric and allowing for linear fluctuations of spherical metric components $h_{\varphi \varphi}=(\sin{\theta})^2 h_{\theta \theta}$ leads to
\begin{equation}\label{standardJT}
 4 \pi v_2^2 \int \mathrm{d}^2 x\sqrt{-\hat{g}} \phi\left(R-\Lambda_2\right) + 8\pi v_2^2 \int_{\partial M} du \,\phi\, K\, ,
\end{equation}
where $\phi$ is identified with $h_{\theta \theta}$ and Dirichlet conditions are set for $\phi$.
It also should be mentioned that the effective two-dimensional cosmological constant $\Lambda_2$ is a combination of the magnetic part of the field strength of \eqref{4dkineticterms} with background value \eqref{backgroundfieldstrengths} and \eqref{gaugescalarFI}. In (\ref{standardJT}) we have also added a boundary term originating from dimensionally reducing a Gibbons-Hawking-York term \cite{Nayak:2018qej}. In the following boundary terms will not be included.\\

\noindent Now we add fluctuations for the gauge fields in the gravity and vector multiplet along the $U(1)$ under which $u$ and $v$ are charged. Spherical symmetry is respected by setting the $\varphi$ and $\theta$ components in the corresponding combination of gauge fields to zero.
The resulting vector field provides an effective photon. 
Assuming $A_{\hat{\mu}}^\Lambda = A^\Lambda A_{\hat{\mu}}$ with $\hat{\mu} \in \left\{ t,r\right\}$ and $A^\Lambda$ denoting a constant direction within the two $U(1)$'s gives the two-dimensional kinetic term
\begin{equation}
    \frac{1}{g_2 ^2}\int d^2 x \sqrt{-\hat{g}}\left( 1 + 2 \phi\right) F_{\hat{\mu}\hat{\nu}} F^{\hat{\mu}\hat{\nu}} ,
\end{equation}
where $F_{\hat{\mu}\hat{\nu}}$ is the fieldstrength of $A_{\hat{\mu}}$ and
\begin{equation}\label{eq:u1fluc}
\frac{1}{g_2^2} = 4 \pi\,v_2 ^2 A^\Lambda I_{\Lambda\Sigma}A^{\Sigma}
\end{equation}
with $\mathcal{I}_{\Lambda \Sigma}$ being the imaginary part of the period matrix depending on the horizon values of the gauge scalars $z$. The fluctuating gauge field $A_{\hat{\mu}}$ will be called photon in the rest of the paper. \\

\noindent
The gravitinos have to fulfill specific $\gamma_{01}$ projections.
The projection conditions on the hyperinos (with solutions (\ref{projections})) induce projection conditions on the supercurrent. 
Since the supercurrent acts as the source of the gravitini, this further induces projection conditions on the gravitini via their equations of motion. From an s-wave perspective these projections are needed to lose the angular spin connection components and also to obtain two-dimensional mass terms. Furthermore, when working out all contributions to the gravitino terms of \eqref{4dkineticterms}, one encounters couplings of $\psi_r$ to $\psi_t$. These are not compatible with unbroken two dimensinal ${\cal N}=\left( 2,2\right)$ supersymmetry. When applying the correct projection these kinds of terms vanish as we will see below. We must also apply a spherical projection on the gravitini in order to emulate the spherical symmetry, linking $J^{\varphi}_A$ to $J^{\theta}_A$\,. To be more precise, when expressing the supercurrents explicitly via the matter sector as in \eqref{4dsupercurrentcoupling} spherical symmetry manifests as 
\begin{equation}\label{eq:jphi}
 J_{A}^{\theta}=-\sin{\theta}\gamma_{23}J_{A}^{\varphi}\,.
\end{equation}
This should also be respected by the gravitino sector. All in all, we apply
\begin{equation}
\label{gravitinoprojections}
\psi_{r/t A}
=\gamma_{01}{(\sigma^3)_A}^{B}\psi_{r/t B}\,,\,\,\,
\psi_{\theta A}
=-\gamma_{01}{(\sigma^3)_A}^{B}\psi_{\theta B}\,,\,\,\,
\psi_{\varphi A}
= \sin{\theta}\,\gamma_{23}\,\psi_{\theta A}\,.
\end{equation}
Let us first understand the general structure of the gravitino contribution to \eqref{4dkineticterms} while only applying the $\gamma_{23}$ projection of \eqref{gravitinoprojections}. First, the field strength with its background value \eqref{backgroundfieldstrengths} effectively acts as a mass term  because it can be rewritten via the first equation of \eqref{BPSlinearcombi} as a purely geometric term; the same is also true for the mass matrix contribution. This can be expressed geometrically via the second equation of \eqref{BPSlinearcombi}. We assume no angular dependence of the gravitino components $\psi_{tA}, \psi_{rA}$ and $\psi_{\theta A}$  ($\psi_{\varphi A}$ is fixed by \eqref{gravitinoprojections}).
Hence, we get the following expression for the kinetic terms of the gravitini 
\begin{align}
    & S \supset \int \mathrm{d}^4 x \; e_{\varphi}^{3}\bar{\psi}_{t}^A \left(-2\gamma_3 \partial_r \psi_{\theta A}+\frac{\text{i}}{ r}\sigma_{ A B}^{3}\gamma^{0 2}\psi_{\theta}^B   +\frac{\text{i} v_2}{2 }\left(-\psi_r^{B}\epsilon_{AB}+\sigma_{ A B}^{3}\gamma^{01}\psi_r^{B}\right)\right)\nonumber \\
     +&\int \mathrm{d}^4 x \;  e_{\varphi}^{3}\bar{\psi}_{r}^A\left( 2\gamma_3 \partial_t \psi_{\theta A}+\frac{r}{v_1^2}\left(-\gamma_{013}\psi_{\theta A}+\text{i} \sigma_{ AB}^{3}\gamma_{12}\psi_{\theta}^{B}\right)+\frac{\text{i}v_2}{2 }\left(\psi_{ t}^{B}\epsilon_{A B}-\sigma_{ AB}^{3}\gamma^{01}\psi_t^{B}\right)     \right)\nonumber \\
      +& 2\int \mathrm{d}^4 x \; e_{\varphi}^{3}\bar{\psi}_{\theta}^A\left(\frac{v_1}{ r v_2}\gamma_{123}\partial_t \psi_{\theta A}-\frac{r}{v_1 v_2}\gamma_{023}\partial_{r}\psi_{\theta A}-\gamma_{3}\partial_t \psi_{r A}+\gamma_{3}\partial_{r}\psi_{t A}\right.\nonumber \\ 
      &+\frac{1}{2 v_1 v_2}\left(-\gamma_{023}\psi_{\theta  A}-\gamma_{23}\psi^B_{\theta}\epsilon_{AB}-\text{i}\sigma_{ AB}^{3}\psi_{\theta}^B\right)\nonumber\\
      &\left. + \frac{r }{2 v_1^2 }\left(\gamma_{013}\psi_{r A}+\text{i}\sigma_{ AB}^{3}\gamma^{21}\psi_r^{B}\right)+\frac{\text{i} }{2 r}\sigma_{ AB}^{3}\gamma^{20}\psi_t^{B}    \right)\,, \label{gravitinoaction}
\end{align}
where so far only the relationship of $\psi_{\varphi A}$ to $\psi_{\theta A}$ \eqref{gravitinoprojections} has been used.
Now we observe couplings of $\psi_t$ to $\psi_r$, couplings of $\psi_{\theta}$ to $\psi_t$ and $\psi_r$ and also terms exclusively consisting of $\psi_{\theta}$. As there is no kinetic term for the gravitini in two dimensions, any consistent dimensional reduction should exclude couplings of $\psi_t$ to $\psi_r$. A close inspection of these terms in the first two lines of \eqref{gravitinoaction} shows that these terms vanish when applying \eqref{gravitinoprojections}.\\
After use of the $\gamma_{01}$ projections in \eqref{gravitinoaction} the $\psi_{\theta}$ to $\psi_{\theta}$ couplings can be brought into the form,
\begin{equation}\label{thetakineticterm}
8 \pi \int \mathrm{d}^2 x \epsilon^{2 3 \hat{\nu}\hat{\mu}}\left(\bar{\psi}_{\theta}^{A}\gamma_{\nu}\gamma_{2 3}\nabla_{\mu}\psi_{\theta A} \right) \,.
\end{equation}
In the purely gravitational sector a linearized approximation is used to arrive at the Jackiw-Teitelboim action, where the dilaton appears only as a Lagrange multiplier. This procedure can be emulated here by denoting $\psi_{\theta}$ as the dilatino mode. This immediately implies that the quadratic term \eqref{thetakineticterm} is to be neglected. In accordance, after applying the $\gamma_{01}$ projections, the remaining terms of \eqref{gravitinoaction} consist purely of $\psi_{t/r A}$ couplings to $\psi_{\theta A}$. 
To be specific, we solve projection conditions (\ref{gravitinoprojections}) explicitly and replace four component spinors by a single Grassmann field according to 
\begin{equation}\label{eq:progra}
\begin{array}{c c}
\psi_{t/r1} \rightarrow \left(\begin{array}{c} 1 \\ \text{i}\\-\text{i}\\-1\end{array}\right) \psi_{t/r1} , & 
\psi_{t/r2} \rightarrow \left(\begin{array}{c} 1 \\ \text{i}\\\text{i}\\1\end{array}\right) \psi_{t/r2} , \\
\psi_{\theta 1} \rightarrow \left( \begin{array}{c} 1 \\ \text{i}\\ \text{i}\\1 \end{array}\right) \psi_{\theta 1} ,&
\psi_{\theta  2}\rightarrow \left( \begin{array}{c} 1 \\ \text{i}\\ -\text{i}\\-1 \end{array}\right) \psi_{\theta 2} .\end{array}
\end{equation}
We focus on all terms in which $\psi_{t/r A}$ and $\psi_{\theta A}$ mix and perform the spherical integration. In addition, we partially integrate those terms in which derivatives of  $\psi_{\theta A}$ appear. In the resulting expression $\psi_{\theta A}$ solely represents a fermionic Lagrange multiplier.
We just give the final answer in two-dimensional conventions ($z=t+y$, $\bar{z}= t-y$, $y= v_1^2/r$)
\begin{align} \label{Gravitinidimreduced}
32\pi v_2\int \mathrm{d}z\mathrm{d}\bar{z} &\left( \psi_{\theta 1}
\left(\nabla_{\bar z}\psi_{z 1}^{\ast}-\nabla_{z}\psi_{\bar{z}1}^{\ast}+ \frac{  \text{i} }{2 y}\psi_{z 2}\right) \right. \nonumber\\
& \left. +\psi_{\theta 2} \left( \nabla_{z}\psi_{\bar{z}2}^{\ast} -\nabla_{\bar z}\psi_{z 2}^{\ast}-\frac{ \text{i} }{2 y}\psi_{ \bar{z} 1}\right)+\text{h.c.}  \right)\nonumber\\
=32 \pi v_2^2\int \mathrm{d}z\mathrm{d}\bar{z} &\left(\lambda_{ 1} \left(  \nabla_{\bar z}\psi_{z 1}^{\ast}-\nabla_{z}\psi_{\bar{z}1}^{\ast}+ \frac{  \text{i} }{2 y}\psi_{z 2}\right)  \right.\nonumber\\
&\left. + \lambda_{ 2} \left(  \nabla_{z}\psi_{\bar{z}2}^{\ast} -\nabla_{\bar z}\psi_{z 2}^{\ast}-\frac{ \text{i} }{2 y}\psi_{ \bar{z} 1}\right)+\text{h.c.}  \right)\,,
\end{align}
where the covariant derivatives are given by
$$\nabla_z \psi_{\bar{z},A} =\partial_z \psi_{\bar{z},A}
  +{\left(\sigma^3\right)_A}^B \frac{1}{4y} \psi_{\bar{z},B}\,\,\, ,\,\,\,
  \nabla_{\bar{z}} \psi_{z,A} =\partial_{\bar{z}} \psi_{z,A}
  +{\left(\sigma^3\right)_A}^B  \frac{1}{4y} \psi_{z,B}
$$
and in the last two lines we have introduced the dilatino with
$\psi_{\theta A}=e_{\theta}^{2}\lambda_A$.\\ 

\noindent
It should be noted that this procedure is quite natural from a supersymmetric perspective. Recall that for the gravitational sector the role of the dilaton was played by $h_{\theta \theta}$. Together with $\psi_{\theta}$ these should constitute the dilaton multiplet. So far however the degrees of freedom do not match up. Whereas the dilaton which naturally appears as the metric fluctuation $h_{\theta \theta}$ must be real, we have double the amount of degrees of freedom for the dilatino. In ${\cal N} = (2,2)$ JT gravity there are two dilaton multiplets \cite{Forste:2017apw}. Here however we do not consider the full Kaluza-Klein reduction, which would furnish the $U(1)_A$ field strength accompanied by the missing bosonic degree of freedom in the dilaton multiplet. Hence we must set reality conditions on the dilatino.\\
Setting $\lambda_1=\text{i}\lambda_2*$, we arrive at
\begin{equation}
32 \pi v_2^2\int \mathrm{d}z\mathrm{d}\bar{z}\lambda_{ 1}\; \left(\nabla_{\bar{z}} \left( \psi_{z1}^* +\text{i} \psi_{z2}\right)
-\nabla_z\left( \psi_{\bar{z}1}^* +\text{i} \psi_{\bar{z}2}\right)  -\frac{1}{2 y}\left( \psi_{\bar{z}1} + \text{i} \psi_{\bar{z} 2}\right) + \text{h.c.} \right) .
\end{equation}
The conditions for the dilatino modes are also applied when calculating the four-point function \eqref{eq:rest}.

\subsubsection{Matter Sector}\label{mattersector}
Now we consider the kinetic and mass terms of the hypermultiplet in the near horizon limit. These match the corresponding terms for the twisted chiral and anti-chiral multiplets in section \ref{EquationsofMotionandCurrents}.\\
The relevant terms of the $\mathcal{N}=2$ Lagrangian are, the kinetic terms for our matter fields
\begin{equation}\label{4dkinetic}
    \int \mathrm{d}^4 x \sqrt{-g}\left( \rule{0pt}{2.5ex}h_{a b}\nabla_{\mu}q^{a}\nabla^{\mu}q^{b}-\text{i}  \left(\bar{\zeta}^{\alpha}\gamma^{\mu}\nabla_{\mu}\zeta_{\alpha}-\text{h.c.}\right)
    \right) \,,
\end{equation}
with the general mass terms
\begin{equation}\label{hypermassterms}
   \int \mathrm{d}^4 x \sqrt{-g}   \left(\rule{0pt}{2.5ex}g^2
   4 h_{a b}k_{\Lambda}^{a}k_{\Sigma}^{b} L^\Lambda \bar{L}^\Sigma +g M^{\alpha \beta}\bar{\zeta}_{\alpha}\zeta_{\beta}\,\right),
\end{equation}
with the hyperino mass matrix defined as
\begin{equation}\label{hyperinomassmatrix}
    M^{\alpha \beta}=-\mathcal{U}_{a}^{\alpha A}\mathcal{U}_{b}^{\beta B}\epsilon_{A B}\partial^{a}k^{b}_{\Lambda}L^{\Lambda} 
\end{equation}
and also a Pauli term
\begin{align}\label{hyperinopauliterm}
     \int \mathrm{d}^4 x \sqrt{-g} F_{\mu \nu}^{-\Lambda}I_{\Lambda \Sigma}\bar{\zeta}_{\alpha}\gamma^{\mu \nu}\zeta_{\gamma}\mathcal{C}^{\alpha \gamma}\,,
\end{align}
which effectively acts as a mass term with the background value \eqref{backgroundfieldstrengths}.
For the vevs we have chosen for the hyperscalars (\ref{eq:hyperbg}) and the choice of moment map (\ref{momentmap}), the scalar mass terms of \eqref{hypermassterms} amount to 
\begin{align}
g^2(4 h_{a b}k_{\Lambda}^{a}k_{\Sigma}^{b})&=g^2 a_{\Lambda}a_{\Sigma}\frac{4}{R} (u^2+v^2)\bar{L}^{\Lambda}L^{\Sigma}\nonumber \\&=(u^2+v^2)U'^2\,,
\end{align}
where for the last step the BPS conditions (\ref{BPS}) were used.
The covariant derivatives are given by
\begin{equation}\label{4dcovariantderivative}
    \nabla_{\mu}q =\left(\partial_{\mu}q+g A_{\mu}^{\Lambda}k_{\Lambda}^{q}\right) \, ,\,\,\,
    \nabla_{\mu}\zeta_{\alpha} =\partial_{\mu}\zeta_{\alpha}-\frac{1}{4}\omega_{\mu}^{m n}\gamma_{m n}\zeta_{\alpha}+\Delta_{\mu \alpha}^{\beta}\zeta_{\beta}\,,
\end{equation}
where for the hyperino covariant derivative we have already applied that the Kähler connection is zero. Furthermore, in our approximation (leaving out higher order interaction terms)
\begin{equation}
    \Delta_{\alpha}^{\beta}\zeta_{\beta}=\mathcal{C}_{\beta \gamma}\Delta^{\alpha \gamma}=\mathcal{C}_{\beta \gamma}\left(g A^{\Lambda}\partial_{a}k^{b}_{\Lambda}\mathcal{U}^{a \alpha A}\mathcal{U}_{b A}^{\beta} \right)\,.
\end{equation}
For both the scalar fields and the fermions the coupling to the gauge fields occurs due to the last term in the covariant derivative. As we are considering electric fluctuations around the magnetic background of the solution, we will also assume  general gauge fields $A_{\hat{\mu}}^{\Lambda}$\,, such that not only terms due to \eqref{eq:vector} will occur. In the s-wave approximation the $A_{\varphi}$ couplings should drop out and only the couplings to $A_{\hat{\mu}}^{\Lambda}$ should appear in two dimensions. We will consider these effects, namely the linearized coupling to supergravity modes in the next section \ref{4dlinearizedcoupling}.\\
For the scalar field one straightforwardly arrives at
\begin{equation}\label{scalarfield2d}
 S \supset  4 \pi v_2^2  \int \mathrm{d}^2 x \sqrt{-\hat g}\frac{1}{R}\left(\partial_{\mu}u\partial^{\mu}u+\partial_{\mu}v\partial^{\mu}v+
 \left(u^2+v^2\right)U'^2\right)\,.
\end{equation}
In the near horizon limit $U^\prime$ is constant and (\ref{scalarfield2d}) resembles a free massive scalar on $AdS_2$. 
This leads to the equations of motion
\begin{equation}\label{scalarfieldeom}
 \frac{v_1^2}{r^2}\partial_t^2 u-\frac{r^2}{v_1^2}\partial_r^2 u-\frac{2 r}{v_1^2}\partial_r u-\frac{2}{v_1}u=0\,,
\end{equation}
and the same with $u$ replaced by $v$.\\
Solving \eqref{scalarfieldeom} at large $r$ leads to the solutions of the form 
\begin{equation}\label{largersolution}
u \sim r^{-\Delta_{\pm}}
\end{equation}
with
\begin{equation}
    \Delta_{\pm}=\frac{1\pm 3}{2}\,,
\end{equation}
and the same for $v$.
According to the AdS/CFT dictionary \cite{Witten:1998qj} $u$ and $v$ are dual to conformal primaries of dimension $\Delta_+$ of the emergent $CFT_1$.
For the comparison with the two-dimensional results of section \ref{SupersymmetricJT} it should also be noted that the scalar fields $u,v$ always appear in complex linear combinations with the vevs chosen for the hypermultiplet sector, such that it is convenient to introduce the combinations
\begin{equation}
\label{hypermultipletlinearcombi}
     f:= u-\text{i} v\,,\,\,\, 
 \bar f:= u+\text{i} v\,. 
 \end{equation}
The action for the complex scalars is 
\begin{equation}\label{scalarfield2dc}
S\supset  4 \pi v_2^2  \int \mathrm{d}^2 x \sqrt{-\hat g}\frac{1}{R}\left(\partial_{\mu}f\partial^{\mu}\bar f+\left( f\bar{f}\right)U'^2\right)\,.
\end{equation}
For the hyperinos we impose the projections which led us to (\ref{eq:progra}) earlier. Then terms in the Lagrangian including angular components of the spin connection drop out.
To be more explicit, \eqref{4dkinetic} includes terms of the form
\begin{equation}
    -\text{i}\left(\bar{\zeta}^{\alpha}\gamma^{\theta}\nabla_{\theta}\zeta_{\alpha}+\bar{\zeta}^{\alpha}\gamma^{\varphi}\nabla_{\varphi}\zeta_{\alpha}\right)\,,
\end{equation}
which are set to zero since the $\gamma_{0 1}$ projections commute with $\gamma_{23}$.\\
By use of $M^{1 2}=-\text{i} a_{\Lambda}L^{\Lambda}$, the background value of the field strength (\ref{backgroundfieldstrengths}) and the BPS equations, \eqref{hypermassterms} and \eqref{hyperinopauliterm} combine to a single effective mass term such that one ends up with the following two-dimensional Lagrangian for the hyperinos
\begin{equation}\label{fermions2d}
 S\supset 4 \pi v_2^2  \int \mathrm{d}^2 x \sqrt{-\hat g}\left(-\text{i}\bar{\zeta}^{\alpha}\gamma^{\hat{\mu}}\nabla_{\hat{\mu}}\zeta_{\alpha}+ U'\bar{\zeta}_{1}\zeta_{2}+\text{h.c.}\right)  \,.
\end{equation}
Replacing $\zeta^\alpha$ by the one component fields as in 
(\ref{projections}) and taking variational derivatives of \eqref{fermions2d} gives the following fermionic equations of motion
\begin{equation}
    \label{fermionequations}
  \partial_t \zeta^{1}-\frac{r^2}{v_1^2} \partial_r\zeta^{1}-\frac{r}{2 v_1^2}\zeta^{1}-\frac{\text{i} r}{ v_1^2}\zeta_{2}=0\,,\,\,\,
  \partial_t \zeta^{2}+\frac{r^2}{v_1^2} \partial_r\zeta^{2}+\frac{r}{2 v_1^2}\zeta^{2}+\frac{\text{i} r}{ v_1^2}\zeta_{1}=0\,.
\end{equation}
These describe free massive fermions on AdS$_2$. For large $r$ we arrive at a solution of the form \eqref{largersolution} with 
\begin{equation}
    \Delta_{\pm}=\frac{1\pm 2}{2}\,.
\end{equation}

\subsubsection{Linearized Coupling}\label{4dlinearizedcoupling}
So far we have discussed the dimensional reduction of the fields of the supergravity theory itself, namely metric, photon and gravitinos and also of the matter on this specific background. In this section we want to dimensionally reduce the coupling of the photon and gravitino fluctuations to the matter. To be more explicit, we now perform the dimensional reduction of the part of the action, which quite heuristically may be written as
\begin{equation} \label{eq:gratinicoup}
S \supset \int d^4x \sqrt{-g}\left( h_{\mu\nu} T^{\mu\nu} + \bar{\psi}_\mu ^A J^\mu _A + \bar{J}_\mu ^A \psi^\mu + A_\mu j^\mu _A\right) .
\end{equation}
As explained previously, due to covariance metric fluctuations are already coupled to matter fields, such that we will only mention this schematically.
We have to allow for metric fluctuations in the spherical directions $h_{\varphi\varphi} = \sin^2\theta\, h_{\theta \theta}$ and metric fluctuations in the $AdS_2$ direction $h_{\hat{\mu}\hat{\nu}}$ in \eqref{4dkinetic}, to arrive at a structure like the first term in (\ref{eq:gratinicoup}). Now integrating over the spherical directions gives the two-dimensional action \cite{Nayak:2018qej}
\begin{equation}
    \begin{aligned}\label{eq:linearc}
 -4 \pi v_2^2 \int \mathrm{d}^2 x \sqrt{- \hat g}\left( h_{\hat \mu \hat \nu}T^{\hat \mu \hat \nu}+2\phi T^{\theta \theta}\right)\,,
 \end{aligned}
\end{equation}
where the metric fluctuation $h_{\theta \theta}$ has been identified with the dilaton $\phi$. The effective 2d energy-momentum conservation reads
\begin{equation}\label{eq:emconDR}
v_1 ^4 \partial_t T_{tt} - r^2 \partial_r\left( r^2 T_{tr}\right) =0 \, , \,\,\,
v_1 ^4 \partial_t T_{tr} - r\partial_r\left( r^3 T_{rr}\right) - rv_1^4T_{tt} =0 .
\end{equation}
\\
The coupling of matter fields to the considered $U(1)$ fluctuations (\ref{eq:u1fluc}) are contained in the covariant derivatives in \eqref{4dkinetic} via \eqref{4dcovariantderivative}.\\
Starting from \eqref{4dkinetic} we arrive at 
\begin{align} 
S &\supset -4 \pi v_2 ^2 g\int d^2 x \sqrt{-\hat{g}}\left( \bar{\zeta}^1 \gamma^\mu \zeta_1 -\bar{\zeta}^2 \gamma^\mu \zeta_2 - 2 u\partial^\mu v + 2 v\partial^\mu u\right) a_\Lambda A^\Lambda _\mu \nonumber \\
& = -q_2 \int d^2x \sqrt{-\hat{g}} \left( \bar{\zeta}^1 \gamma^{\hat{\mu}} \zeta_1 -\bar{\zeta}^2 \gamma^{\hat{\mu}} \zeta_2 - 2 u\partial^{\hat{\mu}} v + 2 v\partial^{\hat{\mu}} u\right)  A _{\hat{\mu}} 
\label{4delectriccoupling}
\end{align}
where the charge $q_2$ is
\begin{equation}
q_2 = 4\pi v_2^2 g a_\Lambda A^\Lambda .
\end{equation}
Notice also that the sum over $\mu \in\left\{ t,r,\theta,\varphi\right\}$ is reduced to $\hat{\mu} \in \left\{ t,r\right\}$ because $u$ and $v$ are taken to depend only on $t$ and $r$ and the $\zeta_A$'s are eigenspinors of $\gamma_{01}$ (which anticommutes with $\gamma_{02}$ and $\gamma_{03}$). 
\\
We now express the four component spinors in \eqref{4delectriccoupling} by their one component projections according to (\ref{projections}) and combine two real scalars into one complex scalar \eqref{hypermultipletlinearcombi}. Then (\ref{4delectriccoupling}) equals
 \begin{equation}
\begin{aligned}
 & q_2\int \mathrm{d}^2 x \sqrt{-\hat g}\left(-4 \text{i} A_{t}e^{t}_{0}(\zeta_{1}^{\ast}\zeta_1 -\zeta_{2}^{\ast}\zeta_{2})-4 \text{i} A_{r}e^{r}_{1}(-\zeta_{1}^{\ast}\zeta_1 -\zeta_{2}^{\ast}\zeta_2)    \right)\\
 +& q_2\int \mathrm{d}^2 x \sqrt{- \hat g}\left(-\text{i}  A_{\hat \mu}\,\partial^{\hat \mu}\,\bar{f}\,f+\text{i}  A_{\hat \mu}\,\partial^{\hat \mu}f \,\bar{f} \right)\,.
\end{aligned}
\end{equation}
Taking the variational derivative of \eqref{4delectriccoupling} with respect to $A_{\hat{\mu}}$ we arrive at the currents
\begin{equation}\label{eq:emcur}
\begin{aligned}
j_t&=-\text{i} q_2\left(f\partial_{t}\bar{f} -\bar{f}\partial_{t}f\right)+4 \text{i} q_2\frac{r}{v_1}\left(-\zeta_1^{\ast}\zeta_1+\zeta_{2}^{\ast}\zeta_{2}\right)\,,\\
j_r&=-\text{i} q_2\left(f\partial_{r}\bar{f} -\bar{\chi}\partial_{r}f\right)+4 \text{i} q_2\frac{v_1}{r}\left(-\zeta_1^{\ast}\zeta_1-\zeta_{2}^{\ast}\zeta_{2}
\right)\, .
\end{aligned}
\end{equation}
The current conservation equation is given by
\begin{equation}\label{eq:abelconNH}
    \partial_{t}j^t+\partial_{r}j^r=0\,.
\end{equation}
The coupling of gravitino to supercurrent can directly be read off from the general $\mathcal{N}=2$ supergravity Lagrangian \cite{n=2sugra}
\begin{equation}
\label{4dsupercurrentcoupling}
 \int \mathrm{d}^4 x \sqrt{-g}\left(-2U_{a A \alpha}\nabla^{\mu}q^{a}\bar{\psi}_{\mu}^{A}\zeta^{\alpha}-2U_{a A \alpha}\nabla_{\mu}q^{a}\bar{\psi}_{\nu}^{A}\gamma^{\mu \nu} \zeta_{\alpha}-2 g i N_{A}^{\alpha}\bar{\psi}_{\mu}^{A}\gamma^{\mu}\zeta_{\alpha}    \right)\,,
\end{equation}
with 
\begin{equation}
    N_{\alpha}^{A}=2 \mathcal{U}_{\alpha a}^{A}k^{a}_{\Lambda}\bar{L}^{\Lambda}\,.
\end{equation}
For clarity we will give the individual supercurrents of \eqref{4dsupercurrentcoupling} explicitly before moving to two-dimensional conventions. We will use the linear combinations \eqref{hypermultipletlinearcombi} and explicit solutions to  spinor projection conditions as in (\ref{projections}) and (\ref{eq:progra}). Then we get one component supercurrents (for a relation to four component spinors see (\ref{eq:procur})),
\begin{equation}
\begin{aligned}
  J_1^{t}&=\frac{\sqrt{2}}{\sqrt{R}}(\frac{v_1^2}{r^2}\partial_t \chi+\partial_r\chi)\zeta^{1}-\frac{\text{i}}{r}\chi \zeta_{2}\,,\,\,\,
   J_2^{t}=\frac{\sqrt{2}}{\sqrt{R}}(-\frac{v_1^2}{r^2}\partial_t \bar \chi+\partial_r \bar \chi)\zeta^{2}-\frac{\text{i}}{r}\bar \chi \zeta_{1}\,,\\
     J_1^{r}&=\frac{\sqrt{2}}{\sqrt{R}}(-\partial_t \chi-\frac{r^2}{v_1^2}\partial_r\chi)\zeta^{1}-\text{i} r\chi \zeta_{2}\,,\,\,\,
  J_2^{r}=\frac{\sqrt{2}}{\sqrt{R}}(-\partial_t \bar \chi+\frac{r^2}{v_1^2}\partial_r\bar\chi)\zeta^{2}+\text{i}i r\bar \chi \zeta_{1}\,\,.
  \end{aligned}
\end{equation}
  The supercurrent components satisfy conservation equations 
  \begin{align}
   \partial_t J_1^{t}+\partial_r J_1^{r}-\frac{ r}{2 v_1^2}\left( J_1^t-\text{i} J_2^{t \ast}\right) +\frac{\text{i} }{2 r}J_2^{ r \ast}&=0\,,\label{4dsupercurrentconservation1}\\
   \partial_t J_2^{t}+\partial_r J_2^{r}+\frac{ r}{2 v_1^2}\left( J_2^t-\text{i}J_1^{t \ast}\right)-\frac{\text{i}}{2 r}J_1^{ r \ast}&=0\,.
   \label{4dsupercurrentconservation2}
\end{align}
  For the angular components the following relation holds
  \begin{equation}
      J_A^{\theta}=-\sin{\theta}\gamma_{23}J_A^{\varphi}\,.
  \end{equation}
  The explicit form is
  \begin{equation}
J_1^{\theta}=\frac{\sqrt{2}}{\sqrt{R}}\left(\frac{v_1}{r b}\partial_t f-\frac{r}{v_1 b}\partial_r f\right)\zeta^{1}+\frac{\text{i}}{b} f \zeta_{2},\,\,\,
    J_2^{\theta}=\frac{\sqrt{2}}{\sqrt{R}}\left(\frac{v_1}{r b}\partial_t f+\frac{r}{v_1 b}\partial_r f\right)\zeta^{2}-\frac{\text{i}}{b}f \zeta_{1}\,.
\end{equation}
  Now we want to perform  the dimensional reduction of the linearized supercurrent gravitino coupling which in terms of four component spinors reads
  \begin{equation}
\begin{aligned}\label{linearizedsupercurrentcoupling}
   4 \pi v_2^2\int \mathrm{d}^2 x \sqrt{-\hat g}\left(\bar \psi^{A}_{\hat \mu}J^{\hat \mu}_A+ 2\bar{\psi}_{\theta}^{A}J^{\theta}_A+ \text{h.c.}    \right)\,.
   \end{aligned}
\end{equation}
  Plugging in our explicit solutions to projection conditions in terms of one component Grassmann fields (see (\ref{projections}), (\ref{eq:progra}) and (\ref{eq:procur})) yields
  \begin{equation}\label{linearizedgravitinicurrentcoupling}
  16 \pi v_2^2 \int \text{d}^2 x \left( \psi_{1\hat{\mu}}^* J_1^{\hat{\mu}} -\psi_{2\hat{\mu}}^* J_2^{\hat{\mu}} -
  2\psi_{1\theta}^* J_1^{\theta} + 2\psi_{2\theta}^* J_2^{\theta}+ \text{c.c.} \right) .
  \end{equation}
  with $\hat{\mu} \in \left\{ r, t\right\}$. The $\psi_{A\hat{\mu}}$ are the gravitini of the $\left(2,2\right)$ Sugra multiplet in 2d whereas $\psi_{\theta A} $ are the superpartners of the dilaton, the dilatini.
  
\subsection{Four-Point Function \label{sec:4pt4d}}
In this subsection we will compute four point functions in the spirit of \cite{Liu:1998ty}. We will turn on fluctuations in half of the hypermultiplet as described in section \ref{sec:hyperfluc}. This will backreact and create fluctuations for gravitons, gravitini and photons. 
By expressing these in terms of the hypermultiplet fluctuations and
plugging that into the action one obtains terms which are quartic in
the hypermultiplet fluctuations. According to the AdS/CFT dictionary
those generate fourpoint functions in the dual CFT$_3$. In a certain
limit these match fourpoint functions in the SYK model as was shown
for gravitons and gauge fields in  \cite{Nayak:2018qej},
\cite{Moitra:2018jqs}. This is because the region in which the
geometry differs from $AdS_2 \times S^2$ contributes, in the
considered low frequency approximation, only contact terms
which will be neglected.
In the dual CFT contact terms can be cancelled by local counterterms.
In our case, the calculation corresponding to integrating out gravitons and photons is quite close to the one reported above. For this reason, we can be brief there. For the gravitini, the discussion will be more complicated and, indeed, we will be able only to match a subsector of the SYK result. 

\subsubsection{Integrating Out Gravitons}

We consider metric fluctuations around our solution, impose spherical symmetry and fix diffeomorphisms, i.e.\ we consider \cite{Nayak:2018qej},
 \begin{equation}
 ds^2 = U^2\left(r\right)\left( 1 + h_{tt}\left( r,t\right) \right) dt^2 -U^{-2}\left( r\right) dr^2 -b^2\left( r\right) \left( 1 + h_{\theta\theta}\right) \left( d\theta^2 + \sin ^2 \theta d\varphi^2\right)  ,
 \end{equation}
 where $U\left( r\right)$ and $b\left( r\right)$ are solutions to the BPS equations (\ref{BPS}) ensuring also that they solve Einstein's equations for the given background. 
 For the computation of non-contact contributions to the fourpoint function only the region of spacetime in which the geometry can be nearly approximated by $AdS_2 \times S^2$ is relevant.
 Here, `nearly' means that $U\left( r\right)$ is taken to its Maldacena limit
 \begin{equation}
 U\left( r\right) = \frac{r}{v_1}
 \end{equation}
 where $r$ has been shifted by $r_h$: 
 \begin{equation}
 r - r_h \to r .
 \end{equation}
 For the $S^2$ radius linear deviations from a constant are taken into account 
 \begin{align}
 b\left( r\right) & = v_2 + r , \\
 f\left( b\right) & = f\left( v_2\right) + r f^\prime\left( v_2\right)  \label{eq:expand}, 
 \end{align}
 where in the expansion (\ref{eq:expand}) only the leading contribution is kept for each term. If no further derivatives w.r.t.\ $r$
 are considered this is the limit given in \cite{Nayak:2018qej}
 (adopted to our notation)
 \begin{equation}\label{eq:indlimit}
 U\left( r\right) =\frac{r}{v_1},\,\,\, b\left( r\right) = v_2 ,\, \,\, b^\prime\left( r \right) =1 . 
 \end{equation}
 After integrating over the $S^2$ the relevant part of the on-shell action is as in (\ref{eq:linearc}) with $\phi = h_{\theta\theta}$ and the metric fluctuations are to be replaced by the corresponding solutions of the linearised Einstein equations. This works in the same way as in \cite{Nayak:2018qej}. The explicit form of the energy-momentum tensor does not matter, here. In our case it will have contributions from two real scalars $u$ and $v$ and also from the hyperinos satisfying the projection condition (\ref{projections}). What matters later is that $T^{\hat{\mu}\hat{\nu}}$ with $\hat{\mu },\hat{\nu} \in \left\{0,1\right\}$ matches the result from the twisted chiral multiplet to be discussed in section \ref{Mattercoupled2dsugra}. It is perhaps worthwhile to point out that $T_{\theta\theta}$ is expressed in terms of $T_{\hat{\mu}\hat{\nu}}$ with $\hat{\mu} ,\hat{\nu} \in \left\{0,1\right\}$ by means of energy-momentum conservation. For this it is important to take the nearly $AdS_2\times S^2$ limit (with $b^\prime\left( r\right) =1$) because in the original Maldacena limit $T_{\theta\theta}$ would decouple from the conservation law which would just be the two dimensional energy-momentum conservation. 
 
\noindent  The result for this sector of the on-shell action can be just copied from \cite{Nayak:2018qej} (up to differences in the signature)
 \begin{equation}\label{eq:gravos}
 S_{g,\text{os}} \sim \int dt d r \left( \frac{r^2 v_2 ^3}{v_1 ^2} T_{rr}\frac{1}{\partial_t} T_{tr}  + \frac{v_2 ^3 r}{v_1^2} T_{tr}\frac{1}{\partial_t^2} T_{tr}\right) .
 \end{equation}
To explicitly compute the four point function one should employ
holographic renormalization (for a review see
\cite{Skenderis:2002wp}). For our purpose of comparing to results from
integrating out super Schwarzian modes the regularised version
($r<\infty$) suffices.

\subsubsection{Integrating Out Gaugefields}

According to our discussion after (\ref{4delectriccoupling}) the angular components of the electromagnetic current vanish. 
Spherical symmetry imposes vanishing angular dependence on the gauge fields and we consider only fluctuations for $A_t$ and $A_r$. %
\begin{equation}\label{eq:gfluc}
\partial_{\hat{\mu}}F^{\hat{\mu}\hat{\nu}} = g_2^2 j^{\hat{\mu}} ,
\end{equation}
where $j^{\hat{\mu}}$ is given in terms of hypermultiplet fluctuations (\ref{eq:emcur}). Following \cite{Moitra:2018jqs} we gauge fix $A_t = 0$ and solve (\ref{eq:gfluc}) by
\begin{equation}
A_r = g_2^2 \, \frac{r}{v_1} \partial_t^{-2} j_r 
\end{equation}
resulting in the on-shell action
\begin{equation}\label{eq:abelon}
S_{A,\text{os}} \sim \int dr dt \left(\frac{r}{v_1}\right)^4 j_r \partial_t ^{-2} j_r .
\end{equation}
\subsubsection{Integrating Out Gravitini\label{sec:gravitini}}

Integrating out the gravitini reflects some new aspects and we will be more detailed, here. 
We gauge fix
\begin{equation}
\psi_{t A}  = 0 ,\,\,\, A \in \left\{ 1,2\right\}.
\end{equation}
From spherical symmetry we deduced that $J^\varphi_A$ is related to $J^\theta_A$ (\ref{eq:jphi}). This is compatible with the gravitini equations if we impose
the last condition in (\ref{gravitinoprojections}).
At the moment no near horizon limit is considered.
Then the remaining gravitini equations of motion are from the variation of (\ref{4dkineticterms}) plus (\ref{eq:gratinicoup}) with respect to the gravitini
\begin{align}
J_1^t = & -\frac{2}{b} \gamma_3 \partial_r \psi_{\theta 1} - \frac{b^\prime}{b^2} \gamma_3 \psi_{\theta 1} + \frac{b^\prime U}{b} \gamma_{123} \psi_{r1} + \text{i}\left( \frac{U^\prime}{Ub} +\frac{b^\prime}{b^2}\right) \gamma_{02} \psi_{\theta 2}^* +\nonumber\\& + \frac{\text{i}}{2}\left[ \frac{Ub^\prime}{b} - U^\prime + \left(  \frac{Ub^\prime}{b} + U^\prime \right) \gamma_{01}\right] \psi_{r2}^* \\
J_2^t  = & -\frac{2}{b} \gamma_3 \partial_r \psi_{\theta 2} - \frac{b^\prime}{b^2} \gamma_3 \psi_{\theta 2} + \frac{b^\prime U}{b} \gamma_{123} \psi_{r 2} +\text{i}\left( \frac{U^\prime}{Ub} +\frac{b^\prime}{b^2}\right) \gamma_{02} \psi_{\theta 1}^*+\nonumber \\ &+ \frac{\text{i}}{2}\left[ U^\prime -\frac{Ub^\prime}{b} + \left(  \frac{Ub^\prime}{b} + U^\prime \right) \gamma_{01}\right] \psi_{r1}^* \\
J_1 ^r = & \frac{2}{b} \gamma_3 \partial_t \psi_{\theta 1} - \left( \frac{ U U^\prime}{b} + \frac{ U^2 b^\prime}{b^2}\right) \gamma_{013} \psi_{\theta 1} -\text{i} \left( \frac{ U U^\prime}{b} + \frac{ U^2 b^\prime}{b^2}\right) \gamma_{12}\psi_{\theta 2}^* \\
J_2 ^r = & \frac{2}{b} \gamma_3 \partial_t \psi_{\theta 2} - \left( \frac{ U U^\prime}{b} + \frac{ U^2 b^\prime}{b^2}\right) \gamma_{013} \psi_{\theta 2} -\text{i} \left( \frac{ U U^\prime}{b} + \frac{ U^2 b^\prime}{b^2}\right) \gamma_{12}\psi_{\theta 1}^* \\
J_1^\theta = & -\frac{1}{b}\gamma_3 \partial_t \psi_{r1} +\left(\frac{U U^\prime}{2b} +\frac{ U^2 b^\prime}{2 b^2}\right)\gamma_{013}\,\psi_{r1} +\frac{1}{Ub^2}\gamma_{123} \partial_t \psi_{\theta 1} -\frac{U^\prime}{2b^2} \gamma_{023}\,\psi_{\theta 1} - \nonumber \\ & -\frac{U}{b^2}\gamma_{023}\partial_r \psi_{\theta 1} +\left[ \left( \frac{U b^\prime}{2b^3} -\frac{U^\prime}{2b^2}\right)\gamma_{23} +\text{i} \left( \frac{U b^\prime}{2b^3} +\frac{U^\prime}{2b^2}\right)\right]\psi_{\theta 2}^* + \nonumber \\ &+
\text{i}\left( \frac{U U^\prime}{2b} +\frac{ U^2 b^\prime}{2 b^2}\right) \gamma_{12} \psi_{r2}^*\\
J_2^\theta = & -\frac{1}{b}\gamma_3 \partial_t \psi_{r2} +\left(\frac{U U^\prime}{2b} +\frac{ U^2 b^\prime}{2 b^2}\right)\gamma_{013}\,\psi_{r2} +\frac{1}{Ub^2}\gamma_{123} \partial_t \psi_{\theta 2} -\frac{U^\prime}{2b^2} \gamma_{023}\,\psi_{\theta 2} - \nonumber \\ & -\frac{U}{b^2}\gamma_{023}\partial_r \psi_{\theta 2} -\left[ \left( \frac{U b^\prime}{2b^3} -\frac{U^\prime}{2b^2}\right)\gamma_{23} -\text{i} \left( \frac{U b^\prime}{2b^3} +\frac{U^\prime}{2b^2}\right)\right]\psi_{\theta 1}^* + \nonumber \\ &+
\text{i}\left( \frac{U U^\prime}{2b} +\frac{ U^2 b^\prime}{2 b^2}\right) \gamma_{12} \psi_{r1}^* .
\end{align}

\noindent In the following we will reduce these four component spinor equations each to a one component equation.  In (\ref{projections}) we had reduced the four component hyperinos to one component by imposing that the frozen scalars $R$ and $D$ do not change under susy transformation (together with chirality). This amounts to a reduction of the supercurrents (other projections will vanish) which we denote as
\begin{equation}\label{eq:procur}
\begin{array}{c c}
J_1 ^{t/r} \rightarrow \left(\begin{array}{r} 1 \\ -\text{i}\\-\text{i}\\1\end{array}\right) J_1 ^{t/r} , & 
J_2 ^{t/r} \rightarrow \left(\begin{array}{r} 1 \\ -\text{i}\\\text{i}\\-1\end{array}\right) J_2 ^{t/r} , \\
J_1^\theta \rightarrow \left( \begin{array}{r} 1 \\ -\text{i}\\ \text{i}\\-1 \end{array}\right) J_1^\theta ,&
J_2^\theta \rightarrow \left( \begin{array}{r} 1 \\ -\text{i}\\ -\text{i}\\1 \end{array}\right) J_2^\theta .
\end{array}
\end{equation}
For the gravitini we keep only those components whose equations are sourced by the projected supercurrents. This amounts to imposing projection conditions (\ref{gravitinoprojections}) or explicitly to replace four component spinors by one component ones according to (\ref{eq:progra}).
In the following we will use one component spinors only. For those,
the remaining non trivial gravitini equations read
\begin{align}
J_1 ^t = &\frac{2}{b}\partial_r \psi_{\theta 1}+\frac{b^\prime}{b^2}\psi_{\theta 1} -\frac{b^\prime U}{b}\psi_{r 1} -\text{i}\left( \frac{U^\prime}{Ub} +\frac{b^\prime}{b^2}\right)\psi_{\theta 2}^* +\text{i}\frac{Ub^\prime}{b} \psi_{r 2}^* ,\label{eq:j1t}\\
J_2 ^t = &\frac{2}{b}\partial_r \psi_{\theta 2}+\frac{b^\prime}{b^2}\psi_{\theta 2} +\frac{b^\prime U}{b}\psi_{r 2} -\text{i}\left( \frac{U^\prime}{Ub} +\frac{b^\prime}{b^2}\right)\psi_{\theta 1}^* -\text{i}\frac{Ub^\prime}{b} \psi_{r 1}^* ,\\
J_1 ^r = & -\frac{2}{b}\partial_t \psi_{\theta 1} +\left( \frac{ U U^\prime}{b} + \frac{U^2 b^\prime}{b^2}\right) \psi_{\theta 1} -\text{i} \left( \frac{ U U^\prime}{b} + \frac{U^2 b^\prime}{b^2}\right) \psi_{\theta 2}^* ,\label{eq:j1r}\\
J_2 ^r = & -\frac{2}{b}\partial_t \psi_{\theta 2} -\left( \frac{ U U^\prime}{b} + \frac{U^2 b^\prime}{b^2}\right) \psi_{\theta 2} +\text{i} \left( \frac{ U U^\prime}{b} + \frac{U^2 b^\prime}{b^2}\right) \psi_{\theta 1}^* ,\label{eq:j2r} \\
J_1 ^\theta = & \frac{1}{b}\partial_t \psi_{r 1} +\left(\frac{UU^\prime}{2b} +\frac{U^2b^\prime}{2b^2}\right)\psi_{r 1} + \frac{1}{Ub^2}\partial_t \psi_{\theta 1} -\frac{U^\prime}{2b^2}\psi_{\theta 1} -\frac{U}{b^2}\partial_r \psi_{\theta 1} +\nonumber \\ & \quad +\text{i} \frac{Ub^\prime}{b^3}\psi_{\theta 2}^* -\frac{\text{i}}{2} \left( \frac{UU^\prime}{b} +\frac{U^2 b^\prime}{b^2}\right)\psi_{r2}^* , \label{eq:j1T}\\
J_2 ^\theta = & \frac{1}{b}\partial_t \psi_{r 2} -\left(\frac{UU^\prime}{2b} +\frac{U^2b^\prime}{2b^2}\right)\psi_{r 2} - \frac{1}{Ub^2}\partial_t \psi_{\theta 2} -\frac{U^\prime}{2b^2}\psi_{\theta 2} -\frac{U}{b^2}\partial_r \psi_{\theta 2} +\nonumber \\ & \quad +\text{i} \frac{Ub^\prime}{b^3}\psi_{\theta 1}^* +\frac{\text{i}}{2} \left( \frac{UU^\prime}{b} +\frac{U^2 b^\prime}{b^2}\right)\psi_{r1}^* .\label{eq:j2T}
\end{align}
From this set of equations one can derive the following conservation laws
\begin{multline}\label{eq:cons1}
\partial_t J_1 ^t +\partial_r J_1 ^r - \frac{U U^\prime}{2}\left( J_1^t -\text{i} J_2 ^{t*}\right)+ \frac{2b^\prime}{b}J_1 ^r -\text{i}\frac{U^\prime}{2U} J_2 ^{r *}+ U b^\prime \left( J_1^\theta - \text{i} J_2 ^{\theta *}\right)   \\ 
= \left( \frac{UU^\prime b^\prime}{b^2}+ \frac{U^2 b^{\prime\prime}}{b^2}+\frac{U U^{\prime\prime}}{b} -\frac{U^2 b^{\prime 2}}{b^3}\right)\left( \psi_{\theta 1} - \text{i}\psi_{\theta 2}^*\right) ,
\end{multline}
\begin{multline}\label{eq:cons2}
\text{i}\left(\partial_t J_2 ^{t*} +\partial_r J_2 ^{r*}\right) - \frac{U U^\prime}{2}\left( J_1^t -\text{i} J_2 ^{t*}\right)-\frac{U^\prime}{2U}J_1 ^r +\text{i}\frac{2b^\prime}{b} J_2 ^{r *}+ U b^\prime \left( J_1^\theta - \text{i} J_2 ^{\theta *}\right)   \\ 
= \left( \frac{UU^\prime b^\prime}{b^2}+ \frac{U^2 b^{\prime\prime}}{b^2}+\frac{U U^{\prime\prime}}{b} -\frac{U^2 b^{\prime 2}}{b^3}\right)\left( \psi_{\theta 1} - \text{i}\psi_{\theta 2}^*\right) .
\end{multline}
Due to the non-vanishing right hand sides in (\ref{eq:cons1}) and (\ref{eq:cons2}) these may not look like proper conservation laws. However, imposing the BPS conditions (\ref{BPS}) it is not difficult to check that the right hand sides vanish. For us there remains a problem, though. In the limit (\ref{eq:indlimit}) the right hand sides do not vanish. This is because the limit is not consistent with the BPS conditions (for consistency one would have to include corrections to $U$, i.e.\ the $AdS_2$ geometry, which we do not want to consider). Another, more technical, problem is that the $J_A ^\theta$ enter only in the combination $J_1 ^\theta -\text{i} J_2 ^{\theta *}$. This means, that one cannot use conservation laws to express the other combination, $J_1 ^\theta +\text{i} J_2 ^{\theta *}$, in terms of $J_A^t$ and $J_A^r$. This does not pose an immediate problem. However, when we will later integrate over the super-Schwarzian modes in section \ref{sec:schwmodes} we will find that the result can be expressed by 2d supercurrents only. The dilatino source corresponding to $J_A^\theta$ will not appear. Both these problems can be addressed by restricting ourselves to a subsector
\begin{equation}\label{eq:rest}
\psi_{\theta 1} -\text{i}\psi_{\theta 2}^* = 0.
\end{equation}
Such a constraint puts the right hand sides of (\ref{eq:cons1}) and (\ref{eq:cons2}) to zero and also solves our second problem since, in the on shell action, $J_1 ^\theta +\text{i} J_2 ^{\theta *}$ couples just to the lhs of (\ref{eq:rest}). From the gravitini equations (\ref{eq:j1r}) and (\ref{eq:j2r}) we learn that (\ref{eq:rest}) constrains
\begin{equation}\label{eq:conrest}
 J_1^r -\text{i}J_2^{r*} =0.
 \end{equation}
After imposing spherical symmetry, the projections (\ref{eq:procur}), (\ref{eq:progra}) and the restriction (\ref{eq:rest}) (implying (\ref{eq:conrest})) the relevant part of the on-shell action takes the form
\begin{multline}\label{eq:osreduced}
  \int \mathrm{d}^4 x \sqrt{- g}\left(\bar{\psi}_{\mu}^{A}J_A^{\mu}+\text{h.c.}\right)\to \\ S_{\psi ,\text{os}} \sim \int dr dt \left[ \left( J_1^r +\text{i}J_2^{r*}\right)^*\left( \psi_{1r} - \text{i}\psi_{2r}^*\right) - 2 \left( J_1^\theta -\text{i} J_2^{\theta *}\right)^* \left( \psi_{1\theta} + \text{i}\psi_{2\theta}^*\right) +\text{c.c.} \right]
\end{multline}
where on the rhs of (\ref{eq:osreduced}) one component fields appear.
This can now be computed along the following steps. First one expresses $J_A^r$ by the gravitini equations (\ref{eq:j1r}) and (\ref{eq:j2r}). The appearing time derivatives of $\psi_{Ar}$ can be expressed by means of gravitini equations (\ref{eq:j1T}), (\ref{eq:j2T}). Taking also the limit (\ref{eq:indlimit}) one arives at
\begin{equation}
S_{\psi ,\text{os}}\sim \int dt dr \left[\left( \psi_{\theta 1} + \text{i}\psi_{\theta 2}^*\right)^* \left( J_1^\theta - \text{i}J_2^{\theta *} + \frac{v_1}{2 r v_2 ^2}\partial_t\left( \psi_{\theta 1} + \text{i}\psi_{\theta 2}^*\right) \right) + \text{c.c.}\right]
\end{equation}
The $\psi_{\theta A}$ can be expressed as solutions to (\ref{eq:j1r}) and (\ref{eq:j2r})
\begin{equation}
\psi_{\theta 1} +\text{i}\psi_{\theta 2}^* = -\frac{v_2}{2} \partial_t^{-1} \left( J_1 ^r +\text{i} J_2 ^{r*}\right) ,
\end{equation}
whereas the combination of $J^{\theta}_A$ can be replaced by means of the conservation laws (\ref{eq:cons1}), (\ref{eq:cons2}). The final result reads
\begin{multline}\label{eq:gravitiniout}
S_{\psi, \text{os}}\sim \int dt dr \left\{ \rule{0pt}{3ex} v_2\left( J_1^r + \text{i}J_2^{r*}\right)^*\right. \times \\ \left.\partial_t ^{-1} \left[ \frac{1}{v_1}\left( J_1^t - \text{i} J_2 ^{t*}\right) -\left(\frac{2v_1}{r}\partial_r+\frac{v_1}{ r^2}\right)\left( J_1^r +\text{i}J_2^{r*}\right) \right] +\text{c.c}\right\}
\end{multline}
where contact terms have been omitted. Notice that, when we go to the near horizon limit (with $b^\prime =0$), the last two terms in (\ref{eq:gravitiniout}) can be expressed via the conservation equations (\ref{4dsupercurrentconservation1}), (\ref{4dsupercurrentconservation2}) as a time derivative of a current component. Therefore, they give rise to contact terms in that limit. The result (\ref{eq:gravitiniout}) will be matched with one obtained by integrating out super-Schwarzian modes in section in \ref{sec:schwmodes}.

\section{Supersymmetric JT}\label{SupersymmetricJT}

In this section we would like to compare our results from the near horizon considerations of the $AdS_4$ supersymmetric black hole to a two dimensional configuration relating supersymmetric JT gravity to the super-Schwarzian effective theory on the boundary. We will consider Euclidean signature and take for the $AdS_2$ metric
\begin{equation}
ds^2 = \frac{dz d\bar{z}}{y^2} = \frac{ dx^2 + dy^2}{y^2} ,\,\,\, z= x+ \text{i} y .
\end{equation}
The coordinate $x$ can be viewed as Euclidean time. For the matter multiplet we will not take directly what we get from dimensional reduction of half the hyper multiplet which we turned on as a probe in the previous section. Instead, we will use two twisted chiral respectively anti-chiral multiplets. They share many features with the probe of the previous section. The major difference is that they are not charged under an extra $U(1)$ but under the $U(1)$ mediated by the 2d graviphoton. This would correspond to a Kaluza Klein $U(1)$ in the dimensional reduction setup. The reason is that integrating out the graviphoton can be directly associated to integrating out a bosonic mode in super reparametrisations of the boundary. The dynamics of this boundary mode is contained in an effective super-Schwarzian action. If instead, we considered the original $U(1)$ gauge field we would need to add an extra phase mode to the boundary as it was done in \cite{Moitra:2018jqs}. This would correspond to a straightforward repetition of the calculation presented in \cite{Moitra:2018jqs}. In the following, we will match results on a qualitative level not taking into account numerical factors. Further we will not identify 2d probe fields with 4d probes but rather present a map between conserved currents. 

\subsection{Minimal $ \mathbf{ 2d\; \mathcal{N}=(2,2)}$ Supergravity\label{subsection:minimalsugra}}
In this section we summarize the Euclidean $ 2d\; \mathcal{N}=(2,2)$ supergravity construction of \cite{Forste:2017apw}.\\
With superspace coordinates
\begin{equation}
    z^\pi=\left(z,\theta^+,\bar \theta^+;\bar z, \theta^-, \bar \theta^-\right)\,
\end{equation}
we have rigid superspace derivatives
\begin{align}
    \partial_z\,,\;\; D_+=\frac{\partial}{\partial
  \theta^+}+\frac{1}{2}\bar \theta^+ \partial_z\,,\; \;\bar
  D_+=\frac{\partial}{\partial \bar \theta^+}+\frac{1}{2}
  \theta^+ \partial_z , \\  
     \partial_{\bar z}\,,\;\; D_-=\frac{\partial}{\partial
  \theta^-}+\frac{1}{2}\bar \theta^- \partial_{\bar z}\,,\;\; \bar
  D_-=\frac{\partial}{\partial \bar \theta^-}+\frac{1}{2}
  \theta^-\partial_{\bar z} ,\;  
\end{align}
which fulfil the anticommutation relations
\begin{equation}
    \left\{D_+,\bar D_+\right\}=\partial_z\,,\;\;\; 
    \left\{D_-,\bar D_-\right\}=\partial_{\bar z}\,.
\end{equation}
In general complex conjugation for fermionic quantities such as the supercharges is given by
\begin{equation}
 \left(Q_+\right)^{*}=   \bar Q_- \,  \, , \;\;\;\;  \left(Q_-\right)^{*}=   \bar Q_+ \,.
\end{equation}
Applying axial torsion constraints and solving them in conformal gauge gives the following supercovariant derivatives
\begin{align}
 \nabla_+&=e^{\bar \sigma}\left(D_+ +\text{i} \left(D_+\sigma\right)
 \bar M\right)\,, \nonumber\\
 \nabla_-&=e^{\bar \sigma}\left(D_- -\text{i} \left(D_-\sigma\right)
 \bar M\right)\,, \nonumber \\
 \bar \nabla_+&=e^{\sigma}\left(\bar D_+ +\text{i} \left(\bar D_+\bar \sigma\right) M\right)\,,\nonumber \\
 \bar \nabla_-&=e^{\sigma}\left(\bar D_- -\text{i} \left(\bar D_-\bar \sigma\right) M\right)\,. \label{supercovariantderivatives}
\end{align}
Here, $\sigma,\bar{\sigma}$ refer to the conformal factors since in $U(1)_A$ supergravity the geometric quantities are given in terms of chiral/anti-chiral fields. $M,\bar{M}$ are convenient linear combinations of the Lorentz and $U(1)_A$ generators.\\
The superconformal factor takes on the following form on the $AdS_2$ geometry,
\begin{align}\label{Explicitsuperconformalfactor}
    \sigma=-\frac{1}{2}\log\left(\frac{1}{2y_c}\right) - \frac{i}{4y_c}\theta^+\theta^-\,,
   \end{align}
   where $y_c$ refers to the chiral basis. It is important to note here that the auxiliary field of the gravity multiplet, which appears as the field multipliying the $\theta^+\theta^-$ factor in \eqref{Explicitsuperconformalfactor} takes on a non-zero vev. We will see in section \ref{Mattercoupled2dsugra} that this will furnish the mass of the probe matter.\\
   While the starting point of our considerations will indeed be the superspace described above, it is important to see how the structures of \eqref{supercovariantderivatives} map onto $x$-space quantities as the actual physical calculations will take place there.\\
   The $x$-space covariant derivative, which can be deduced by projecting out superspace coordinates in \eqref{supercovariantderivatives} is of the form
   \begin{align}\label{x-spacederivatives}
       \nabla_{\mu}=\partial_{\mu}+\mathcal{J} \Omega_{\mu}+\frac{\mathcal{Y}}{2}A_{\mu}\,,
   \end{align}
   where $\mathcal{J},\mathcal{Y}$ refer to the Lorentz and $U(1)_A$ generator respectively and $\Omega_{\mu}$ and $A_{\mu}$ are the spin connection and graviphoton gauge field. Both $\Omega_{\mu}$ and $A_{\mu}$ are of course implied by the bosonic term of the superconformal factor. $\Omega_{\mu}$ is determined by the real part of $\sigma|$, $A_{\mu}$ by the imaginary part of $\sigma|$, where $|$ denotes the projection on the leading component of a multiplet. For the background \eqref{Explicitsuperconformalfactor} the imaginary part of $\sigma|$ is zero, however we must allow for fluctuations later. In section \ref{Mattercoupled2dsugra} it will be explained how \eqref{x-spacederivatives} acts on the individual component fields of the matter multiplets.\\
   \subsection{$\mathbf{\mathcal{N}=(2,2)}$ JT Supergravity\label{sec:JTsugra}}
The first term of our two-dimensional action, is the $\mathcal{N}=(2,2)$ JT action, which leaving out the topological term, is given by \cite{Forste:2017apw}:
\begin{equation}\label{supersymmetricJTaction}
  S=
   -\frac{1}{16 \pi G_N}\left[\int\limits_\mathcal{M}
     \mathrm{d}^2 z \mathrm{d}^2 \theta \mathcal{E}^{-1}\Phi \left(
     R+2\right) +\text{h.c.}
  +2\int\limits_{\partial\mathcal{M}} \mathrm{d}u
     \mathrm{d}\vartheta \mathrm{d}\bar \vartheta \left(\Phi_b+\bar \Phi_b\right)
     \mathcal{K} \right]\,.
\end{equation}
$\mathcal{E}^{-1}$ refers to the chiral density, which is needed to correctly define chiral superspace integration, $R$ to the chiral supercurvature and $\mathcal{K}$ to the extrinsic supercurvature. Furthermore, the dilaton naturally also appears as a chiral and anti-chiral field with the field content $\Phi \supset \varphi,\lambda_{\alpha},B$ and $\Phi \supset \bar{\varphi},\bar{\lambda}_{\alpha},\bar{B}$ .\\
We should also think about what \eqref{supersymmetricJTaction} implies in $x$-space and how it can be related to the four-dimensional model of the previous section. 
We repeat the analysis of \cite{Forste:2017apw} for the bosonic fields: the variations with respect to the supergravity auxiliary fields fix the dilaton auxiliary fields to be related to the dynamical bosonic field of the dilaton, such that one ends up with the following bosonic part of the JT action in $x$-space
\begin{equation}\label{xspaceJTaction}
S_{\text{JT,bos.}}=\frac{\text{i}}{16 \pi G_N} \int \mathrm{d}z \mathrm{d}\bar{z}\sqrt{g}\left(\varphi\left(\mathcal{R}+ \text{i} \mathcal{F}+2\right)+\bar{\varphi}\left(\mathcal{ R}- \text{i} \mathcal{ F}+2\right)   \rule{0pt}{2.5ex}  \right)\,.
\end{equation}
Recall, that the supersymmetric JT action not only furnishes a  dynamical term for metric fluctuations, it also allows the gravitino to acquire a kinetic term, as the standard gravitino term vanishes in two dimensions. In superconformal gauge the gravitino appears as the fermionic components of the conformal factor.
The coupling of dilatino to gravitino is
\begin{align}
S_{\text{JT,ferm.}}= & \frac{1}{2 \pi G_N}\int \mathrm{d}z\mathrm{d}\bar{z}\left[\lambda_+\left( \nabla_{\bar{z}}\psi_{\bar{+}z} -\nabla_z
\psi_{\bar{+}\bar{z}} +\frac{\text{i}}{2y}\psi_{-\bar{z}}\right)\right. + \nonumber \\
&+\left. \lambda_-\left( \nabla_z\psi_{\bar{-}\bar{z}} -\nabla_{\bar{z}} \psi_{\bar{-}z} -\frac{\text{i}}{2y} \psi_{+z}\right) + \bar{\lambda}_+\left( \nabla_z\psi_{+\bar{z}}-\nabla_{\bar{z}} \psi_{+z}+\frac{\text{i} }{2y}\psi_{\bar{-}} z\right)\right. + \nonumber \\
& + \left. \bar{\lambda}_- \left( \nabla_{\bar{z}} \psi_{-z} - \nabla_z \psi_{-\bar{z}} - \frac{\text{i} }{2y}\psi_{\bar{+} z}\right)\right]\, .
\label{xspacedilatinogravitinocoupling1}
\end{align}
 Assuming real curvature constraints in \eqref{xspaceJTaction} implies $\varphi=\bar{\varphi}$ and hence a real Lagrange multiplier coupled to the Ricci scalar, which is just the canonical form of the JT action. In order for dilaton degrees of freedom to match, reality constraints have to be applied to the dilatino modes.
 We apply Majorana conditions, which in our conventions amount to $\lambda_{+}=\bar \lambda_{+}$ and $\lambda_{-}=\bar \lambda_{-}$. This results in 
\begin{align}
S_{\text{JT,ferm.}}=&\frac{1}{4 \pi G_N}\int \mathrm{d}z\mathrm{d}\bar{z}\,\left[\lambda_{+}\left( \nabla_{\bar z}\psi_{\bar{+} z }-\nabla_{z}\psi_{\bar{+}\bar{z}}+       \frac{  \text{i} }{2 y}\psi_{-\bar{z} }+\nabla_{ z}\psi_{+ \bar z }-\nabla_{ \bar z}\psi_{+ z }+       \frac{\text{i}}{2 y}\psi_{\bar{-} z} \right)\right.\nonumber \\
&+\left.\,\lambda_{-}\left( \nabla_{ z}\psi_{\bar{-}\bar z }-\nabla_{\bar z}\psi_{\bar{-}z }-         \frac{  \text{i} }{2 y}\psi_{+ z}+\nabla_{ \bar z}\psi_{-  z }-\nabla_{ \bar z}\psi_{- \bar  z }-\frac{\text{i}}{2 y}\psi_{\bar{+} z}\right)\right]\,\,.
\label{xspacedilatinogravitinocoupling2}
\end{align}
\subsubsection{Graviphoton Kinetic Term}
As we have introduced a gauge field in the covariant derivatives and will treat gauged matter below, we should also add a kinetic term for the gauge field. First consider the supersymmetric Gauss-Bonnet term
\begin{equation}
  S=
   -\frac{1}{16 \pi G_N}\int\limits_\mathcal{M}
     \mathrm{d}^2 z \mathrm{d}^2 \theta \mathcal{E}^{-1} 
     R +\text{h.c.} \,.
\end{equation}
Here, when moving to $x$-space, the field strength drops out and one recovers the standard Gauss-Bonnet term. Therefore a further term is required in order for a kinetic term for the gauge field to appear,
\begin{equation}
  S=
   -\frac{1}{2 \pi G_N}\int\limits_\mathcal{M}
     \mathrm{d}^2 z \mathrm{d}^2 \theta  \mathrm{d}^2 \bar\theta 
     R \, \bar R \,.
\end{equation}
Integrating to $x$-space gives 
\begin{align}\label{2dfieldstrengthkinetic}
  S=
   -\frac{\text{i}}{4 \pi G_N}\int\limits_\mathcal{M}
     \mathrm{d}^2 z \sqrt{g} F_{z \bar z}F^{z \bar z},
\end{align}
where the curvature $\mathcal{R}$ drops out.
Here, we should specify exactly how the superconformal factor is related to the graviphoton.\\
In general $\text{Im}(\sigma)|$ constitutes the gauge field in Lorentz gauge $A_{\mu}=\epsilon_{\mu \nu}\partial^{\nu}\text{Im}(\sigma)|$.\\
Hence,
\begin{equation}
 \partial_z \,\text{Im}(\sigma)|=-\frac{1}{2} A_{z}\,,\,\,\,
 \partial_z \text{Im}(\bar \sigma)|=\frac{1}{2} A_{z}\,,
\end{equation}
and also
\begin{equation}\begin{aligned}
 \partial \bar{\partial}\text{Im}(\sigma)|& =\frac{1}{4}F_{z \bar z}\,.
\end{aligned}\end{equation}
From a two-dimensional perspective the term  \eqref{2dfieldstrengthkinetic} will reduce to the kinetic term of the internal $U(1)_A$ mode at the boundary and from a four-dimensional perspective this term corresponds to the Kaluza-Klein field strength.\\
\subsubsection{JT Supergravity and the Super-Schwarzian}
Just as delineated in \cite{Maldacena:2016upp} integrating out the dilaton as a Lagrange multiplier, constrains the geometry to $AdS_2$, while at the same time reducing the action to an integral over the boundary, which in our supersymmetric case is \cite{Forste:2017apw}
\begin{equation}
    S_\text{eff}=  -\frac{1 }{8 \pi
  G_N}\int\limits_{\partial\mathcal{M}} \mathrm{d}u
  \mathrm{d}\vartheta\mathrm{d}\bar \vartheta  (\Phi_b+\bar
  \Phi_b)\mathcal{K}\,. \label{eq::effectiveboundaryaction1} 
\end{equation}
Setting Dirichlet conditions for the dilaton and calculating the supercurvature then leaves us with the explicit form for the effective action of the system:
\begin{equation}
    S_\text{eff}=-\frac{1 }{2 \pi G_N}
  \int\limits_{\partial\mathcal{M}} \mathrm{d}u
  \mathrm{d}\vartheta\mathrm{d}\bar \vartheta \, \varphi_b\,
  \text{Schw}\left(t,\xi, \bar \xi ; u, \vartheta, \bar \vartheta
  \right)\,,
\label{eq:effective}
\end{equation}
where $\varphi_b$ is the boundary value for the leading component of $\Phi$ and $ \text{Schw}\left(t,\xi, \bar \xi ; u, \vartheta, \bar \vartheta
  \right)$ refers to the $\mathcal{N}=2$ super-Schwarzian,
   which is defined by
\begin{equation}\label{superschwarzian}
       \text{Schw}\left( x,\xi, \bar \xi ;
  u, \vartheta, \bar \vartheta \right) =  \frac{( D_{\bar\vartheta} \bar\xi^\prime)}
{ D_{\bar\vartheta} \bar \xi}- \frac{( D_{\vartheta} \xi^\prime)}{ D_{\vartheta}  \xi}-2
\frac{\xi^\prime
\bar \xi^\prime}{(D_{\vartheta}\xi)( D_{\bar\vartheta} \bar \xi)} \,,
\end{equation}
with $\xi,\bar{\xi}$ general super-reparametrisations of the boundary and $u,\vartheta,\bar{\vartheta}$ the boundary superspace coordinates. The super-reparametrisations are subject to superconformal constraints \eqref{Superreps}.\\ Whereas for the bosonic case the Schwarzian action describes the soft mode of reparametrisations of time, here the situation is generalised to superspace. The super-Schwarzian constitutes the effective action of reparametrisations of time, and the fermionic coordinates of the boundary superspace.\\ 
   It should be mentioned that the $x$-space expressions for the kinetic terms of the gravity multiplet given in the previous sections could also be reduced to  boundary expressions individually. Here, we have assumed the super-Schwarzian as the boundary effective action due to the arguments presented in \cite{Forste:2017apw} and then projected down to $x$-space. Alternatively, it should in principle also be possible to perform everything entirely in superspace.
  
\subsection{Matter Coupled to $\mathbf{2d\; \mathcal{N}=(2,2)}$ Supergravity \label{Mattercoupled2dsugra}}
Now we also want to add supersymmetric matter to the JT supergravity theory. This is done by straightforwardly adding a matter term, such that the matter field only couples to the metric and not the dilaton. The field can then  be considered to be moving on a pure $AdS_2$ geometry, such that the usual $AdS/CFT$ dictionary can be applied.\\
Hence, we must only work out what the coupling of the superconformal factor to a locally supersymmetric matter multiplet in superspace amounts to in components in $x$-space.
\subsubsection{Chiral vs.\ Twisted Chiral}
 For global supersymmetry, there are two main ways to build symmetric theories: setting chiral constraints or setting twisted chiral constraints on a general superfield. While the former is charged under $U(1)_{V}$ and uncharged under $U(1)_{A}$, the opposite is true for the latter. Since we are interested in constructing matter gauged under the graviphoton of the supercurvature multiplet, which can be related to a bosonic mode in super-reparameterisations of the boundary, we must set twisted chiral constraints given by
\begin{equation}\label{Equation:twistedchiralrigid}
    \Bar{D}_{+}\chi=0\,,\;\;\;\;D_{-}\chi=0\,.
\end{equation}
Here a crucial difference arises to the Lorentzian case \cite{Hull:2008de}. Whereas complex conjugation of \eqref{Equation:twistedchiralrigid}  implies the conditions for the associated twisted anti-chiral field for Lorentzian singature, here, due to the complex conjugation properties elucidated in section \ref{subsection:minimalsugra} we obtain the same constraints on the complex conjugated field, such that the usual kinetic action would vanish. This implies that we have to choose $\chi$ to be real.\\
\subsubsection{Supersymmetric Action}
For our analysis we must construct superfields which are covariantly twisted chiral, which means that they fulfil the generalisation of \eqref{Equation:twistedchiralrigid} to curved space. Such that a covariantly twisted chiral field is given by
\begin{equation}\label{Equation:twistedchiralcovariant1}
    \Bar{\nabla}_{+}\chi_{\text{cov}}=0\,,\;\;\;\;\nabla_{-}\chi_{\text{cov}}=0\,,
\end{equation}
and a covariantly twisted anti-chiral field by
\begin{equation}\label{Equation:twistedchiralcovariant2}
    \Bar{\nabla}_{-}\Bar{\chi}_{\text{cov}}=0\,,\;\;\;\;\nabla_{+}\Bar{\chi}_{\text{cov}}=0\,.
\end{equation}
Note that the notation $\Bar{\chi}$ here does not refer to complex conjugation. 
The solution of the constraints \eqref{Equation:twistedchiralcovariant1} depends on the charge of the superfield, which we choose to be
\begin{equation}\label{superspacecharge}
    [M,\chi]=-\text{i}\chi\,,\,\,\,
    [M,\bar\chi] =\text{i}\bar\chi\,.
\end{equation}
In superconformal gauge we arrive at the following expressions for the definition of our covariant fields 
\begin{equation}\label{Equation:solutiontwistedcovariant}
    \chi_{\text{cov.}} =e^{-\sigma-\bar{\sigma}}\chi\,,\,\,\,
    \bar{\chi}_{\text{cov.}}=e^{-\sigma-\bar{\sigma}}\bar{\chi}\,.
\end{equation}
The formal expression for the D-term is
\begin{equation}\label{Equation:sugraaction}
    \int \mathrm{d}z\mathrm{d}\Bar{z}\mathrm{d}^2 \theta\mathrm{d}^2\Bar{\theta}\text{E}^{-1} \chi_{\text{cov}}\,\Bar{\chi}_{\text{cov}}\,.
\end{equation}
As we will see below, there will be no need to add another probe term.
\subsubsection{$\mathbf{X}$-Space}
We also have to define how the physical fields, which will appear in $x$-space after superspace integration are defined.\\
The most convenient way to do this is by the projection method:
\begin{align}\label{Equation:projectionmethod}
\left. \chi_{\text{cov}}\right|&= f \;\;\;\;\;\;\;\;\;\quad\quad\quad\quad\quad\quad\;\,\left.\Bar{\chi}_{\text{cov}}\right|=\Bar{f}\\
\nabla_{+}\chi_{\text{cov}}&=\zeta_{+}\;\;\;\;\;\quad\quad\quad\quad\quad\quad\nabla_{-}\chi_{\text{cov}}=\zeta_{-}\\
\bar\nabla_{-}\chi_{\text{cov}}&=\bar\zeta_{-}\;\;\;\;\quad\quad\quad\quad\quad\quad\;\overline{\nabla}_{+}\Bar{\chi}_{\text{cov}}=\Bar{\zeta}_{+}\\
\frac{1}{2} [\nabla_{+},\bar{\nabla}_{-}]\chi_{\text{cov}}&=F\;\;\;\quad\quad\quad\;\;\;\,\;\frac{1}{2} [\bar\nabla_{+}\nabla_{-}]\bar\chi_{\text{cov}}=\bar F\,,
\end{align}
We should also translate the superspace charge \eqref{superspacecharge} into $U(1)_{A}$ and Lorentz charges in $x$-space for the individual component fields \eqref{Equation:projectionmethod}. \\
For the bosonic components
\begin{align}
    [\mathcal{Y},f]=2\;\;\;\;[\mathcal{Y},\bar{f}]=2\,,
\end{align}
and for the fermionic fields
\begin{align}
    [\mathcal{Y},\zeta_{-}/\zeta_{\bar{+}}]=\zeta_{-}/\zeta_{\bar{+}}\,.
\end{align}
For the Lorentz charge we naturally get
\begin{align}
    [\mathcal{J},\bar{\zeta}_{+}/\zeta_{+}]&=-\frac{\text{i}}{2}\bar{\zeta}_{+}/\zeta_{+}\,\\
    [\mathcal{J},\bar{\zeta}_{-}/\zeta_{-}]&=\frac{\text{i}}{2}\bar{\zeta}_{-}/\zeta_{-}\,.
\end{align}
This determines \eqref{x-spacederivatives}. In conformal gauge for \eqref{Explicitsuperconformalfactor} we have $\Omega_{z/ \bar{z}}=\frac{1}{2 y}$.\\

\subsubsection{Breaking Superconformal Symmetry}
We can now just take the formal expression \eqref{Equation:sugraaction} and perform the integration via the chiral density method and then project onto the physical $x$-space fields via \eqref{Equation:projectionmethod}. So far it would seem that as we have only included a D-term, we still have to add an F-term in order to add masses to the fields and break conformal symmetry. However, as we will see, the gauging itself breaks superconformal symmetry. This is due to a theorem first noted in \cite{Gates:1995du}: \\\\
\textit{In 2D, if the spin-0 field of a matter supermultiplet carries a non-trivial realization of an internal symmetry charge that is gauged by a spin-1 field in the superconformal multiplet, the action for the spin-0 field is neither conformally nor superconformally invariant.}\\\\
For the specific case at hand this occurs because in the component expansion of \eqref{Equation:sugraaction} the matter fields couple to the supergravity auxiliary field of \eqref{Explicitsuperconformalfactor}. Therefore the masses are determined by the curvature itself.
\subsection{Equations of Motion and Currents}\label{EquationsofMotionandCurrents}
Now performing the integration of \eqref{Equation:sugraaction} in superspace and then using \eqref{Equation:projectionmethod} we arrive at the two-dimensional matter action for the probe multiplet
\begin{align}
    \frac{\text{i}}{2} \int  \mathrm{d}z\mathrm{d}\Bar{z}&\left[ \left(\partial f\right)\left(\bar{\partial}\bar{f}\right)+\left(\bar{\partial}f\right)\left(\partial \bar{f}\right) +\frac{1}{y^2}f\bar{f}-\frac{1}{2 y}\bar{\zeta}_{-}\partial \zeta_{-}-\bar{\zeta}_{+}\bar{\partial}\zeta_{+}-\text{i} \frac{1}{2 y^2}\bar{\zeta}_{-}\bar{\zeta}_{+}+\right.\nonumber
    \\
    &+f A_{z}\bar{\partial}\bar{f}+f A_{\bar{z}}\partial \bar{f}+\frac{3}{2 y} A_{z}\zeta_{-}\bar{\zeta}_{+}+ \nonumber\\
    &+\left.\bar{\psi}^{+ z}\left(\bar{\zeta}_{+}\partial \bar{f}\right)-\text{i}\bar{\psi}^{+ \bar{z}}\left(e_{\bar{z}}^{\bar{l}}\bar{\zeta}_{-} \bar{f}\right)+\bar{\psi}^{- \bar{z}}\left(\bar{\zeta}_{-}\bar{\partial}f\right)+\text{i}\bar{\psi}^{- z}\left(e_{z}^{l}\zeta_{+}f\right)+\text{h.c.}\right]\, .
\label{xspaceaction2d}
\end{align}
The first line represents the kinetic terms, the second line the linearized coupling to the gauge field and the last line the linearized coupling to the gravitinos.\\
Now focussing on the kinetic terms for a moment we can derive the equations of motion, which take on a simple form by use of \eqref{x-spacederivatives}.\\
For the bosons
\begin{align}
    \partial\bar{\partial}f&=\frac{1}{2 y^2}f\,,\label{bosoneom1}\\
    \partial\bar{\partial}\bar{f}&=\frac{1}{2 y^2}\bar{f}\,,\label{bosoneom2}
\end{align}
and for the fermions
\begin{align}
\nabla_{\bar{z}}\zeta_{+}&=i e^{\bar l}_{\bar z}\bar{\zeta}_{\bar{-}}\,,\label{fermioneom1}\\
    \nabla_{z}\zeta_{-}&=-i e^{l}_{z}\bar{\zeta}_{\bar{+}}\,.
    \label{fermioneom2}
\end{align}
For bosons we can immediately solve the equations asymptotically, which lead to normalizable mode $y^{\Delta_{+}}$ and non-normalizable mode $y^{\Delta_{-}}$ with
\begin{equation}
    \Delta_{\pm}=\frac{1\pm 3}{2}\,.
\end{equation}
Iteration of \eqref{fermioneom1} with the complex conjugate of \eqref{fermioneom1} (and vice versa) provides corresponding asymptotics for fermions with
\begin{equation}
    \Delta_{\pm}=\frac{1\pm 2}{2}\,.
\end{equation}
\subsubsection{Symmetry Currents}
It is now also important for us to write out the symmetry currents of the two-dimensional action. There are currents linked to energy-momentum, $U(1)_A$ and supersymmetry conservation. In principle the currents constitute a multiplet in superspace. More precisely, as we are breaking superconformal symmetry we are essentially introcuding a multiplet of superconformal anomalies, which together with the multiplet of superconformal currents fulfill a conservation equation in superspace. For our purposes we are only interested in $x$-space expressions. A purely superconformal current such as the superstring fulfills the algebraic constraints
\begin{align}\label{superconformalcurrent}
    T^{\mu}_{\mu}=0\;\;\;\;(\gamma^{\mu}S_{\mu})_{\alpha}=0\,,
\end{align}
As we have essentially gauged the tangent space group of the $\mathcal{N}=(2,2)$ superstring and also added massive perturbations, we will have corrections to \eqref{superconformalcurrent}. The energy-momentum tensor acquires a non-zero trace and for the supercurrent, the components $S_{+ \bar{z}},S_{- z}$ and their complex conjugates become non-zero.\\
In a linearized approach we can derive the $U(1)_{A}$ current by taking the variational derivative of \eqref{xspaceaction2d} with respect to the gauge field, which leads to
\begin{equation}\label{U1acurrent}
    j_{z}^{A}=\frac{1}{2}\;\left(\bar f\partial f-f\partial\bar f\right)+\frac{3}{4 y}\zeta_{+}\bar{\zeta}_{+}\,,\,\,\,
    j_{\bar z}^{A}=\frac{1}{2}\left(\bar f\bar\partial f-f\bar\partial\bar f\right)-\frac{3}{4 y}\zeta_{-}\bar{\zeta}_{-}\,,
\end{equation}
where the $A$ denotes the axial nature of these currents.
The conservation equation is given by
\begin{align}\label{eq:abelconJT}
\partial j_{\bar{z}}^{A}+\bar{\partial} j_z^{A}=0\, .   
\end{align}
Due to the internal $U(1)_A$ charge the supercurrents split up into a parts consisting of $f,\zeta_{+},\bar \zeta_{-}$ and a part, which includes $\bar f,\zeta_{-},\bar\zeta_{+}$.\\
The former being
\begin{equation}\begin{aligned}\label{supercurrent}
    S_{\bar{+} \,z}&=\zeta_{+}\partial f\,, \quad \;\;\;\;\;S_{-\,\bar{z}}=\bar{\zeta}_{-}\bar{\partial}f\,,\\
    S_{\bar{+}\, \bar{z}}&=-\text{i} e^{\bar{l}}_{\bar{z}}\bar{\zeta}_{-}f\,, \quad S_{-\,z}=\text{i} e^{l}_{z}\zeta_{+}f \,.
\end{aligned}\end{equation}
and the latter
\begin{equation}\begin{aligned}
S_{+ \,z}&=\bar{\zeta}_{+}\partial\bar{f}\,, \quad \;\;\;\;\;S_{\bar{-}\,\bar{z}}=\zeta_{-}\bar{\partial}\bar{f}\,,\\
    S_{+\, \bar{z}}&=-\text{i} e^{\bar{l}}_{\bar{z}}\bar{\zeta}_{-}\bar{f}\,, \quad S_{\bar{-}\,z}=\text{i} e^{l}_{z}\bar{\zeta}_{+}\bar{f}\,.
\end{aligned}\end{equation}
The conservation equations are
\begin{equation}\label{eq:2dscurcon}\begin{aligned}
 \bar{\partial} S_{\bar{+}z}+\partial S_{\bar{+}\bar{z}}-\frac{\text{i}}{4 y}\left(S_{\bar{+}z}+S_{\bar{+}\bar{z}}\right)+\frac{\text{i}}{2 y}S_{- \bar{z}}&=0\,,\\
  \bar{\partial} S_{\bar{-} z}+\partial S_{\bar{-}\bar{z}}+\frac{\text{i}}{4 y}\left(S_{\bar{-} z}+S_{\bar{-}\bar{z}}\right)-\frac{\text{i}}{2 y}S_{+z}&=0\,.
\end{aligned}\end{equation}
The energy-momentum tensor is most conveniently expressed by splitting it up into fermionic and bosonic contributions. For the bosons we have
\begin{equation}\begin{aligned}\label{bosonicEMT}
    T_{B}^{zz}&=-4y^4\partial f\partial\bar{f}\,,\quad T_{B}^{\bar{z}\bar{z}}=-4y^4\bar{\partial}f\bar{\partial}f\,,\\
    T_{B}^{z\bar{z}}&=2y^2f\bar{f}\,,
\end{aligned}\end{equation}
and for the fermions
\begin{equation}\begin{aligned}\label{fermionEMt}
    T_F^{zz}&=y^3\left(\zeta_{+}\partial\bar{\zeta}_{+}+\bar{\zeta}_{+}\partial\zeta_{+}\right)\,,\quad T_{F}^{\bar{z}\bar{z}}=y^3\left(\zeta_{-}\bar{\partial}\bar{\zeta}_{-}+\bar{\zeta}_{-}\bar{\partial}\zeta_{-}\right)\,,\\
     T_{F}^{z\bar{z}}&=-\text{i} y^2\left(\zeta_{-}\zeta_{+}\right)\,.
\end{aligned}\end{equation}
Now for the combined energy momentum Tensor $T_{\mu \nu}:=T_{B ;\mu \nu}+T_{F ;\mu \nu}$ the conservation equations are
\begin{equation}\label{eq:emconJT}
\partial_{x}T_{xx}+\partial_y T_{x y}=0\,, \,\,\,
\partial_{x}T_{x y}+\partial_{y}T_{y y}+\frac{1}{y}\left(T_{x x}+ T_{y y}\right)=0\,.
\end{equation}

\subsection{Comparison to Four-Dimensional Model\label{sec:4d2sdis}}
We see that at leading order we can let the JT model acquire the same form as the dimensionally reduced model. For the gravity multiplet it is important to note that for the JT term we have a priori already made an assumption by restricting the supercurvature to a real number: $\Phi(R+2)$. This forces a real dilaton and hence also real dilatino structures. In principle one could allow the supercurvature to be a general complex number, which would fix the field strength to a specific value.  In comparison the dimensionally reduced model naturally assumes a real dilaton.\\ Furthermore, for the matter sector we recover
behaviour already noticed in \cite{Nayak:2018qej}. In the dimensionally reduced model, an additional source dilaton coupling $\phi\, T^{\theta \theta}$ appears, which constitutes a deviation from pure JT behaviour as one does not consider matter to dilaton couplings in that approach. Here, we acquire an additional dilatino to supercurrent coupling: $\bar{\psi}_{\theta}^{A}\,S_{A}^{\theta}$. Also, the field strength comes with the standard kinetic term and a linear dilaton coupling, whereas the two-dimensional approach just yields the former.  There is a further deviation related to the different signatures. Recall, that the four-dimensional calculations are performed in Lorentzian signature, whereas the two-dimensional model is Euclidean. In order to add gauged matter, we had to use covariantly twisted chiral and anti-chiral multiplets, which was only possible by applying an additional reality condition. Hence, for two dimensions $f$ and $\bar{f}$ are not linked by complex conjugation and as such are real, whereas for the dimensionally reduced model we had complex bosons. However, the current conservations still match between the two approaches as do the masses.
\subsection{Super-Schwarzian coupled to Matter\label{sec:schwMatt}}
Our general approach will follow the steps outlined in \cite{Nayak:2018qej}. We work out the on-shell action of the two-dimensional probe matter, such that it reduces to a boundary two-point form coupled to fluctuations of the boundary. There are additional complications for the case at hand as the boundary two-point function only takes on a elegant form in superspace. Hence, we work out what the general form of two-point function should be at the boundary due to symmetry restrictions. Then for a general multiplet we allow for superspace fluctuations in this two-point function, and match the resulting $x$-space expressions to the on-shell action of the bulk pulled back to the boundary.\\
Furthermore, we can now calculate the four-point function by combining the previous result with the effective Schwarzian action and integrating out the fluctuations.\\
To recapitulate, we do the following: work out the probe on-shell action implied by \eqref{xspaceaction2d} and reduce to the boundary. This will then be matched with a general result for the boundary two-point form in superspace dictated by symmetry considerations. This superspace action will then in conjunction with \eqref{eq:effective} be used to integrate out the fluctuations.

\subsubsection{On-Shell Action}
Now we want to determine the on-shell action, which will just reduce to a boundary expression. 
A regularised solution to (\ref{bosoneom1}) is
\begin{equation}\label{eq:exactsol}
    f\left(y,\omega\right)=e^{-\left(y-\epsilon\right)\left|\omega\right|}\frac{1+y\left|\omega\right| }{y(1+\epsilon|\omega|)}f\left(\omega\right)\,,
\end{equation}
where a Fourier transform replacing Euclidean time $x$ by $\omega$ has been performed. Solution  (\ref{eq:exactsol}) is unique in that it is regular at $y\to \infty$ and satisfies the boundary condition
$$ f\left(\epsilon ,\omega)\right) = \frac{1}{\epsilon} f\left( \omega\right) $$
for some given $f\left(\omega\right)$. The solution of $\bar{f}\left(y,\omega\right)$ is the same with $f\left(\omega\right)$ replaced by $\bar{f}\left( \omega\right)$. Note, that for more generic masses solution (\ref{eq:exactsol}) is expressed in terms of modified Bessel functions \cite{Freedman:1998tz}. \\

Analogously the solutions to (\ref{fermioneom1}) and (\ref{fermioneom2}) are given by 
\begin{equation}\begin{aligned}\label{fermionsolution}
    \hspace*{-3pt}\zeta_{+}(y,\omega)&=e^{(\epsilon-y)|\omega|}\frac{(1+\omega y +y |\omega|)\zeta(\omega)}{\sqrt{y}(1+\epsilon |\omega|)},\,
    \bar{\zeta}_{-}(y,\omega) =-e^{(\epsilon-y)|\omega|}\frac{(1-\omega y +y |\omega|)\zeta(\omega)}{\sqrt{y}(1+\epsilon |\omega|)}\,\\
\hspace*{-3pt}    \zeta_{-}(y,\omega)&=e^{(\epsilon-y)|\omega|}\frac{(1-\omega y +y |\omega|)\bar{\zeta}(\omega)}{\sqrt{y}(1+\epsilon |\omega|)},\,
    \bar{\zeta}_{+}(y,\omega) =-e^{(\epsilon-y)|\omega|}\frac{(1+\omega y +y |\omega|)\bar{\zeta}(\omega)}{\sqrt{y}(1+\epsilon |\omega|)}\,.\\
\end{aligned}
\end{equation}
\noindent
As can easily be seen from \eqref{fermionsolution}, at the boundary there will only be two fermionic degrees of freedom.
Let us note the explicit boundary behaviour of the solutions above.
The bosonic boundary behaviour is 
\begin{align}\label{bosonsboundaryexpansion}
   f\left(y,\omega\right)\sim f\left(\omega\right)\left(\frac{1}{y}+\frac{\epsilon^2 \omega^2}{2y}-\frac{1}{2}\omega^2 y -\frac{\epsilon^3 \omega^2 |\omega|}{3 y} +\frac{1}{3}\omega^2 y^2  \right) \,,
\end{align}
with the analogous behaviour for $\bar{f}(y,\omega)$.
For the fermions we get
\begin{align}\label{fermionsboundaryexpansion}
    \zeta_{+}\left(y,\omega\right)&\sim  \zeta\left(\omega\right)\left(\frac{1}{\sqrt{y}}+\omega \sqrt{y} +\frac{\epsilon^2 \omega^2}{2 \sqrt{y}}-\frac{1}{2}\omega^2 y^{3/2}-\omega|\omega|y^{3/2}\right)\,,\\
    \zeta_{-}(y,\omega)&\sim  \bar{\zeta}\left(\omega\right)\left(\frac{1}{\sqrt{y}}-\omega \sqrt{y} +\frac{\epsilon^2 \omega^2}{2 \sqrt{y}}-\frac{1}{2}\omega^2 y^{3/2}+\omega|\omega|y^{3/2}\right)\,.
\end{align}
We define the following quantities for which frequency dependence is replaced by dependence on Euclidean time
\begin{equation}\begin{aligned}\label{bosonpositionspacequantities}
     f_{\Delta_{-}}\!\left(x\right)&=\int\mathrm{d} \omega e^{\text{i} \omega x}f\left(\omega\right)\,,\,\,\,
     \bar{f}_{\Delta_{-}}\!\left(x\right) =\int\mathrm{d} \omega e^{\text{i} \omega x}\bar{f}\left(\omega\right)\,,\\
    f_{\Delta_{+}}\!\left(x\right)&=\int\mathrm{d}x' \frac{f_{\Delta_{-}}\!\left( x'\right)}{[x-x']^4} \,,\,\,\,
    \bar{f}_{\Delta_{+}}\!\left( f\right) =\int\mathrm{d}x' \frac{\bar{f}_{\Delta_{-}}\!\left( x'\right)}{[x-x']^4} \,,
\end{aligned}\end{equation}
and similarly for the fermions
\begin{equation}\begin{aligned}\label{fermionpositionspacequantities}
\zeta_{\Delta_{-}}\!\left(x\right)&=\int\mathrm{d} \omega e^{\text{i} \omega x}\zeta\left( \omega\right)\,,\,\,\,
    \bar{\zeta}_{\Delta_{-}}\!\left(x\right) =\int\mathrm{d} \omega e^{\text{i} \omega x}\bar{\zeta}\!\left( \omega\right)\,,\\
  \zeta_{\Delta_{+}}\!\left(x\right)&=\int\mathrm{d}x'\frac{\zeta_{\Delta_{-}}\left( x'\right)}{[x-x']^3}\,,\,\,\, 
  \bar{\zeta}_{\Delta_{+}}\!\left(x\right) =\int\mathrm{d}x'\frac{\bar{\zeta}_{\Delta_{-}}\left( x'\right)}{[x-x']^3}\,.\\
\end{aligned}\end{equation}
In terms of \eqref{bosonpositionspacequantities}, \eqref{fermionpositionspacequantities} the boundary behaviour \eqref{bosonsboundaryexpansion}, \eqref{fermionsboundaryexpansion} amounts to (a dot denotes a derivative w.r.t.\ Eucledian time $x$)
\begin{equation}
  f\left(y,\omega\right)\sim \frac{f_{\Delta_{-}}\left(x\right)}{y}-\frac{2 \epsilon^3 f_{\Delta_{+}}\left(x\right)}{\pi y}+\frac{2 y^2 f_{\Delta_{+}}\left(x\right)}{\pi}-\frac{\epsilon^2 \ddot{f}_{\Delta_{-}}\left(x\right)}{2y}+...
\end{equation}
and for the fermions 
\begin{equation}\begin{aligned}
 \zeta_{+}(y,x)&\sim \frac{\zeta_{\Delta_{-}}(x)}{\sqrt{y}}+\frac{2 \text{i} y^{3/2} \zeta_{\Delta_{+}}(x)}{\pi}-\text{i}\sqrt{y}\dot{\zeta}_{\Delta_{-}}+\frac{2 \epsilon^3 \dot{\zeta}_{\Delta_{+}}}{3 \pi \sqrt{y}}+...\,,\\
  \zeta_{-}(y,x)&\sim \frac{\bar{\zeta}_{\Delta_{-}}(x)}{\sqrt{y}}-\frac{2 \text{i} y^{3/2} \bar{\zeta}_{\Delta_{+}}(x)}{\pi}+\text{i}\sqrt{y}\dot{\bar{\zeta}}_{\Delta_{-}}+\frac{2 \epsilon^3 \dot{\bar{\zeta}}_{\Delta_{+}}}{3 \pi \sqrt{y}}+...\,.
\end{aligned}\end{equation}
\subsubsection{Boundary Super-Space, Two-Point Function}\label{boundarysuperspace}
In superspace, the on-shell action should reduce to the form of a superconformal two-point function. Therefore one must only know what the supertranslation invariant interval on the boundary is and also the structure of chiral or anti-chiral multiplet to give the correct form of the boundary two-point function. The boundary superspace was constructed in the context of the $\mathcal{N}=2$ SYK model \cite{Fu:2016vas}.
The super-derivatives are in our conventions
\begin{equation}
    D_{\vartheta} =\partial_{\vartheta}+\frac{1}{2}\bar{\vartheta}\partial_u\,,\,\,\,
     D_{\bar\vartheta} =\partial_{\bar \vartheta}+\frac{1}{2}\vartheta\partial_u\,,
\end{equation}
with the anticommutation relations
\begin{equation}
    \left\{ D_{\vartheta},\bar D_{\bar\vartheta}\right\}=\partial_u\,.
\end{equation}
Chirality constraints can then be imposed via
\begin{equation}\label{boundaryderivative}
    D_{\vartheta}\bar{\chi}=0\,,\,\,\,
    D_{\bar\vartheta}\chi =0\,.
\end{equation}
Here $\chi,\bar{\chi}$ are general boundary superfields.
The $\mathcal{N}=2$ superreparametrisations 
\begin{equation}
\left( u , \vartheta ,\bar{\vartheta}\right) \to \left( x\left( u, \vartheta, \bar{\vartheta}\right) ,\xi \left( u, \vartheta, \bar{\vartheta}\right), \bar{\xi} \left( u, \vartheta, \bar{\vartheta}\right)\right) ,
\end{equation}
are constrained by
\begin{equation}\label{Superreps}
    D_{\vartheta}\bar\xi =0\,,\;\;\; D_{\vartheta}x\left(u\right)=\frac{1}{2}\bar{\xi}D_{\vartheta}\xi\, ,\;\;\;
    D_{\bar\vartheta}\xi =0\,,\;\;\; D_{\bar\vartheta}x\left(u\right)=\frac{1}{2}\xi D_{\bar\vartheta}\bar\xi\,.
\end{equation}
Their dynamics are effectively described by the super-Schwarzian \eqref{superschwarzian}.
One can solve \eqref{Superreps} for a general component structure ($x$ denotes Euclidean time)
\begin{align}
    x(u)&=u+\epsilon(u)+\frac{1}{2}\left[\vartheta\bar{\eta}(u)+\vartheta\eta(u)\right]\,\nonumber\\
    \xi(u)&=\eta(u)+\vartheta\left[1+\sigma(u)+\frac{1}{2}\Dot{\epsilon}(u)\right]+\frac{1}{2}\vartheta\bar{\vartheta}\Dot{\eta}(u)\,,\label{xspacesuperreps}\\
    \bar \xi(u)&=\bar \eta(u)+\bar\vartheta\left[1-\sigma(u)+\frac{1}{2}\Dot{\epsilon}(u)\right]-\frac{1}{2}\vartheta\bar{\vartheta}\Dot{\bar\eta}(u)\,. \nonumber
\end{align}
We observe that the superreparametrisations can be expressed in $x$-space via four individual modes $\epsilon,\eta,\bar \eta,\sigma$. The first, $\epsilon\left(u\right)$, is the single gravitational mode, which also appears in the purely bosonic setting and as such is the boundary fluctuation, which is linked to the energy-momentum coupling of the bulk on-shell action. In a similar spirit, $\sigma$ represents the boundary degree of freedom of gauge fluctuations $A_{z/ \bar z}$ and $\eta,\bar \eta$ constitute boundary gravitinos and are hence linked to the supercurrent.\\
In order to find the supertranslation invariant boundary superspace structure, we demand the following 
\begin{equation}
    D_{\vartheta}'\Delta_{\text{bdy.}}=D_{\bar{\vartheta}}\Delta_{\text{bdy.}}=0\,.
\end{equation}
The unique solution is \cite{Bulycheva:2018qcp}
\begin{equation}
   \Delta_{\text{bdy.}}=\left[u-u'\right]-\frac{1}{2}\left[\vartheta\bar{\vartheta}+\vartheta'\bar{\vartheta'}+2\bar{\vartheta}\vartheta'\right] \,.
\end{equation}
We can include fluctuations of the boundary super-curve by employing the relations \eqref{xspacesuperreps}
\begin{equation}
   \Delta_{\text{bdy.}}=\left[t(u)-t'(u')\right]-\frac{1}{2}\left[\xi(u)\bar{\xi}(u)+\xi'(u')\bar{\xi'}(u')+2\bar{\xi}(u)\xi'(u')\right] \,.
\end{equation}
We also have to define a boundary multiplet which should have a matter content consistent with the boundary expansions of the bulk matter multiplet. Hence, we define a chiral and an anti-chiral multiplet (with respect to the boundary derivatives). Both will consist of the on-shell boundary degrees of freedom worked out in the previous section. Hence , for the chiral multiplet  we have $\left( \chi_{\text{bdy.}}\supset f_{\Delta_{-}},\zeta_{\Delta_{-}}\right)$ and for the anti-chiral one $\left(\bar{ \chi}_{\text{bdy.}}\supset \bar{f}_{\Delta_{-}},\bar{\zeta}_{\Delta_{-}}\right)$
\begin{align}
    \chi_{\text{bdy.}}\left(u_C\right)&=f_{\Delta_{-}}\left(u_C\right)+\sqrt{2}\vartheta \zeta_{\Delta_{-}}\left(u_C\right)\,,\\
     \bar{\chi}_{\text{bdy.}}\left(u_{AC}\right)&=
     \bar{f}_{\Delta_{-}}\left(u_{AC}\right)-\sqrt{2}\bar \vartheta \bar{\zeta}_{\Delta_{-}}\left(u_{AC}\right)\,.
\end{align}
We end up with the following boundary two-point function coupled to
super-curve fluctuations\footnote{Here, we assume that there is no
  mixing with any other dimension two operator. If there was such a
  mixing we would have to add terms in which one of the $\chi$'s is
  replaced by the corresponding boundary mode. Indeed, there is
  another dimension two operator associated to the dilaton
  \cite{Cvetic:2016eiv}. Since the 2d action does not contain terms
  linear in the twisted multiplet fields and there is no direct
  coupling to the dilaton we do not see how a
  corresponding mixing could arise in an on-shell action.}
\begin{equation}\label{boundaryactionsuperspace}
S_{\chi_{\text{bdy.}}}=\int \mathrm{d}u\mathrm{d}\vartheta\mathrm{d}u'\mathrm{d}\bar \vartheta' \frac{\left[D_{\bar \vartheta}\xi\left(u\right)\right]^3[D_{\vartheta}' \xi'\left(u'\right)]^3}{ \Delta_{\text{bdy.}}^3}\chi_{\text{bdy.}}\left(u\right)\bar{\chi}_{\text{bdy.}}\left(u'\right)\,.
\end{equation}
Plugging in the structure \eqref{xspacesuperreps} and performing the superspace integration of \eqref{boundaryactionsuperspace} will give boundary couplings of the $x$-space matter fields to the fluctuations of \eqref{xspacesuperreps}. These are quite lengthy expressions. Therefore, we give each matter coupling to one of the four fluctuations individually.\\

\noindent\textbf{Matching to Bulk}\\
The internal $\sigma$ mode is coupled in the following way:
\begin{align}\label{internalsigmamodecoupling}
  \hspace{-2.05cm}  \frac{\delta S_{\chi_{\text{bdy.}}}}{\delta \sigma}=6 \left(\bar{f}_{\Delta_{-}}f_{\Delta_{+}}-f_{\Delta_{-}}\bar{f}_{\Delta_{+}}+\zeta_{\Delta_{-}} \bar{\zeta}_{\Delta_{+}}-\bar{\zeta}_{\Delta_{-}}\zeta_{\Delta_{+}}     \right)\,.
\end{align}
The first boundary gravitino mode coupling reads
\begin{equation}\label{boundarygravitinocoupling}
   \hspace{-2.8cm} \frac{\delta S_{\chi_{\text{bdy.}}}}{\delta \eta}=\sqrt{2}\left( 2\bar{f}_{\Delta_{-}} \dot{\bar{\zeta}}_{\Delta_{+}} -3\bar{\chi}_{\Delta_{+}}\bar\zeta_{\Delta_{-}}+3\dot{\bar{f}}_{\Delta_{-}}\bar{\zeta}_{\Delta_{+}}\right)\,,
\end{equation}
and the second is
\begin{equation}
   \hspace{-2.5cm} \frac{\delta S_{\chi_{\text{bdy.}}}}{\delta \bar \eta}=\sqrt{2}\left( -2\chi_{\Delta_{-}} \dot{\zeta}_{\Delta_{+}} +3f_{\Delta_{+}}\zeta_{\Delta_{-}}+3\dot{f}_{\Delta_{-}}\zeta_{\Delta_{+}}\right)\,.
\end{equation}
The boundary graviton couples according to
\begin{align}
    \hspace{-1.35cm} \frac{\delta S_{\chi_{\text{bdy.}}}}{\delta \epsilon}= &-3 \left( f_{\Delta_{-}} \dot{\bar{f}}_{\Delta_{+}}+\bar{f}_{\Delta_{-}} \dot{f}_{\Delta_{+}}+2 f_{\Delta_{+}}\dot{\bar{f}}_{\Delta_{-}}+2\bar{f}_{\Delta_{+}}\dot{f}_{\Delta_{-}}                  \right)\nonumber\\
    &+\left(3\bar{\zeta}_{\Delta_{+}}\dot{\zeta}_{\Delta_{-}}+3 \zeta_{\Delta_{+}}\dot{\bar{\zeta}}_{\Delta_{-}}-\bar{\zeta}_{\Delta_{-}}\dot{\zeta}_{\Delta_{+}}-\zeta_{\Delta_{-}}\dot{\bar \zeta}_{\Delta_{+}}         \right)\,.
\label{boundarygravitoncoupling}
\end{align}
In order to match the expressions with the four-dimensional results we must first express the Schwarzian couplings via the on-shell symmetry currents \eqref{U1acurrent}, \eqref{supercurrent}, \eqref{bosonicEMT}, \eqref{fermionEMt}. The boundary expressions are
\begin{equation}\begin{aligned}\label{boundaryEMT}
 T_{xx}^{\text{bdy.}}&=-\frac{3}{y \pi}\left( f_{\Delta_{-}}\bar{f}_{\Delta_{+}}+\bar{f}_{\Delta_{-}}f_{\Delta_{+}}     \right)\,,\\
 T_{xy}^{\text{bdy.}}&=\frac{1}{\pi}\left(-3f_{\Delta_{+}}\dot{\bar{f}}_{\Delta_{-}}-3\bar{f}_{\Delta_{+}} \dot{f}_{\Delta_{-}}+\bar{\zeta}_{\Delta_{-}}\dot{\zeta}_{\Delta_{+}}+\zeta_{\Delta_{-}}\dot{\bar{\zeta}}_{\Delta_{+}} +\zeta_{\Delta_{+}}\dot{\bar{\zeta}}_{\Delta_{-}}\right)\,,\\
 T_{yy}^{\text{bdy.}}&=\frac{1}{\pi y}\left( 3f_{\Delta_{-}}\bar{f}_{\Delta_{+}}+3\bar{f}_{\Delta_{-}}+2\bar{\zeta}_{\Delta_{-}}\zeta_{\Delta_{+}}+2\zeta_{\Delta_{-}}\bar{\zeta}_{\Delta_{+}}    \right)\,.
\end{aligned}\end{equation}
We can now express \eqref{boundarygravitoncoupling} as
\begin{equation}
   \frac{\delta S_{\chi_{\text{bdy.}}}}{\delta \epsilon}=\pi (T_{t y}-y \partial_t T_{yy})\,.
\end{equation}
The boundary expressions for the supercurrent components are
\begin{equation}\begin{aligned}\label{supercurrentboundary}
 S_{+ z}^{\text{bdy.}}&=\frac{\bar{f}_{\Delta_{-}}\bar{\zeta}_{\Delta_{+}}}{\sqrt{y}\pi}+3\text{i} \frac{\sqrt{y}\bar{\zeta}_{\Delta_{+}}\bar{f}_{\Delta_{-   }}}{\pi}-\text{i}\frac{\sqrt{y}\bar{\zeta}_{\Delta_{+}}\dot{\bar{f}}_{\Delta_{-}}}{\pi}\,,\,\,\,
  S_{\bar{-} z}^{\text{bdy.}}=-\frac{\bar{f}_{\Delta_{-}}\bar{\zeta}_{\Delta_{+}}}{\sqrt{y}\pi}\,,\\
  S_{\bar{-} \bar{z}}^{\text{bdy.}}&=-\frac{\bar{f}_{\Delta_{-}}\bar{\zeta}_{\Delta_{+}}}{\sqrt{y}\pi}+3 \text{i} \frac{\sqrt{y}\bar{\zeta}_{\Delta_{+}}\bar{f}_{\Delta_{+   }}}{\pi}-\text{i}\frac{\sqrt{y}\bar{\zeta}_{\Delta_{+}}\dot{\bar{f}}_{\Delta_{-}}}{\pi}\,,\,\,\,
  S_{+ \bar{z}}^{\text{bdy.}}=\frac{\bar{\chi}_{\Delta_{-}}\bar{\zeta}_{\Delta_{+}}}{\sqrt{y}\pi}\,,\\
S_{\bar{+} z}^{\text{bdy.}}&=-\frac{f_{\Delta_{-}}\zeta_{\Delta_{+}}}{\sqrt{y}\pi}-3 \text{i} \frac{\sqrt{y}\zeta_{\Delta_{-}}\bar{f}_{\Delta_{+   }}}{\pi}+\text{i}\frac{\sqrt{y}\zeta_{\Delta_{+}}\dot{f}_{\Delta_{-}}}{\pi}\,,\,\,\,
  S_{\bar{+} \bar{z}}^{\text{bdy.}} =\frac{f_{\Delta_{-}}\zeta_{\Delta_{+}}}{\sqrt{y}\pi}\,,\\
  S_{- \bar{z}}^{\text{bdy.}} &=\frac{f_{\Delta_{-}}\zeta_{\Delta_{+}}}{\sqrt{y}\pi}-3 \text{i} \frac{\sqrt{y}\zeta_{\Delta_{-}}f_{\Delta_{+   }}}{\pi}+\text{i}\frac{\sqrt{y}\zeta_{\Delta_{+}}\dot{f}_{\Delta_{-}}}{\pi}\,,\,\,\,
  S_{- z}^{\text{bdy.}} =-\frac{f_{\Delta_{-}}\zeta_{\Delta_{+}}}{\sqrt{y}\pi}\,.
\end{aligned}\end{equation}
Equations \eqref{supercurrentboundary} allow us to rewrite \eqref{boundarygravitinocoupling}
\begin{equation}
     \frac{\delta S_{f_{\text{bdy.}}}}{\delta \eta}=\frac{\pi}{\sqrt{2 y}}\left(
     \mbox{\rule{0ex}{2.5ex}}
     \text{i}
     \left(
     \mbox{\rule{0ex}{2ex}}
     S_{+ z}-S_{+ \bar{z}}-S_{\bar{-}z}+S_{\bar{-} \bar{z}}\right)+y \partial_{t}\left(
     \mbox{\rule{0ex}{2ex}}
     S_{+ z}-S_{+ \bar{z}}+S_{\bar{-} z}-S_{\bar{-} \bar{z}}\right)     \right)\,.
\end{equation}
Similarly for $\bar \eta$ we get 
\begin{align}
     \frac{\delta S_{\chi_{\text{bdy.}}}}{\delta \bar \eta}=\,\frac{\pi}{\sqrt{2 y}}\left(
     \mbox{\rule{0ex}{2.5ex}}
     \text{i}\left(
     \mbox{\rule{0ex}{2ex}}
     S_{\bar{+} z}-S_{\bar{+}\bar{z}}-S_{- z}+S_{-\bar{z}}\right)+y \partial_{t}\left(
     \mbox{\rule{0ex}{2ex}}
     S_{\bar{+}z}-S_{\bar{+}\bar{z}}+S_{- z}-S_{- \bar{z}}\right)      \right)\,.
\end{align}
The values of the gauge current at the boundary are 
\begin{equation}\begin{aligned}
 j_{z}^{A}&=\frac{3 \text{i}}{2 \pi }\left( f_{\Delta_{-}}\bar{f}_{\Delta_{+}} -f_{\Delta_{-}}\bar{f}_{\Delta_{+}} +\bar{\zeta}_{\Delta_{-}}\zeta_{\Delta_{+}}-\zeta_{\Delta_{-}}\bar{\zeta}_{\Delta_{+}}        \right)\,,\\
  j_{\bar z}^{A}&=-j_{z}\,.
\end{aligned}\end{equation}
This allows us to express \eqref{internalsigmamodecoupling} as
\begin{equation}
\hspace{-8cm}\frac{\delta S_{\chi_{\text{bdy.}}}}{\delta \sigma}= 2 \pi j_{y}^{A}\,.
\end{equation}
\subsection{Four-Point Function/Integrating out the Fluctuations \label{sec:schwmodes}}
To the action (\ref{boundaryactionsuperspace}) we must also add the super-Schwarzian \eqref{superschwarzian}. This is a quite lengthy expression. Fortunately, we only have to consider expressions at most quadratic in the fields appearing on the rhs of (\ref{xspacesuperreps}), leading to
\begin{equation}\label{superschwarzianquadratic}
    S_{\text{Schw}}=\int\mathrm{d}u\left[-\frac{1}{2}\ddot{\epsilon}\left(u\right)^2+2\sigma\ddot{\sigma}\left(u\right)-4\eta\ddot{\bar{\eta}}\left(u\right)-4\bar{\eta}\ddot{\eta}\left(u\right)\right]\,.
\end{equation}
We now have to work out the equations of motion of the fluctuations in order to integrate them out. We get contributions from the kinetic terms \eqref{superschwarzianquadratic} and from the coupling of these modes to matter, which was worked out in the previous subsection. The general form for the resulting on-shell action is 
\begin{align}
    S=\int\mathrm{d}u &\left[\left(\partial_u^{-4}\frac{\delta S_{\chi_{\text{bdy.}}}}{\delta \epsilon}\right)\left(\frac{\delta S_{\chi_{\text{bdy.}}}}{\delta \epsilon}\right) -\frac{1}{4}\left(\partial_u^{-2}\frac{\delta S_{\chi_{\text{bdy.}}}}{\delta \sigma}\right)\left(\frac{\delta S_{\chi_{\text{bdy.}}}}{\delta \sigma} \right)\right.\nonumber\\ +&\left. \frac{1}{4}\left(  \partial_{u}^{-3}\frac{\delta S_{\chi_{\text{bdy.}}}}{\delta \bar \eta} \right)\left(\frac{\delta S_{\chi_{\text{bdy.}}}}{\delta \eta} \right)+\frac{1}{4}\left(  \partial_{u}^{-3}\frac{\delta S_{\chi_{\text{bdy.}}}}{\delta \eta} \right)\left(\frac{\delta S_{\chi_{\text{bdy.}}}}{\delta \bar\eta} \right)   \right]\,,
\label{eq:bdos}
\end{align}
For comparison with results from the 4d calculation performed in section \ref{sec:4pt4d} we would like to rewrite these expressions in terms of two dimensional integrals. To this end, one inserts a `constructive identity', $\int dy \partial_y$, and employs conservation equations \cite{Nayak:2018qej}. This works well for the first, third and fourth term  in (\ref{eq:bdos})  leading to
\begin{align}
 S_{\epsilon}&=-\frac{\pi^2}{4} \int\mathrm{d}x \mathrm{d}y\, y^2 \left(\partial_{x}^{-2}T_{x x}\right)\left(2 T_{x y}-2 y \partial_{x}T_{yy}\right)\,,\label{eq:2doneps}\\
 S_{\eta \bar{\eta}}&=\frac{\pi^2 \text{i}}{8}\int\mathrm{d}x \mathrm{d}y \left(\partial_{x}^{-1}\left(S_{+ z}+S_{+ \bar{z}}+S_{\bar{-}z}+S_{{\bar{-}\bar{z}}}
 \right)\left(S_{\bar{+}z}-S_{\bar{+}\bar{z}}-S_{- z}+S_{-\bar{}}\right)\right)&\nonumber\\
 &+ \frac{\pi^2 \text{i}}{8}\int\mathrm{d}x \mathrm{d}y \left(\partial_{x}^{-1}\left(S_{\bar{+}z}+S_{\bar{+}\bar{z}}+S_{- z}+S_{- \bar{z}}\right)\left(S_{+ z}-S_{+ \bar{z}}-S_{\bar{-} z}+S_{\bar{-}\bar{z}}\right)\right)\,.\label{eq:2doneta}
\end{align}
For the contribution corresponding to integrating out the gauge field there is a subtlety which has been pointed out in \cite{Moitra:2018jqs}. In our approximation ($\omega y \ll 1$) $j_y ^A$ does not depend on $y$,
and therefore inserting $\int dy \partial_y$ on $j_y^A \partial_x^{-2}j_y^A$  would return zero. The authors of \cite{Moitra:2018jqs} considered just a charged scalar. Their argument is based on the observation that the current contains the scalar and its complex conjugate in an antisymmetrised way. Therefore, only products between different modes in an expansion like (\ref{bosonsboundaryexpansion}) contribute. At the given approximation this includes the first two lowest powers in $y$ yielding a factor of $y$ (since $\Delta_+ + \Delta_- =1$). Another factor of $1/y$ appears due to a $y$ derivative (in the first term on the rhs of (\ref{U1acurrent})). The same arguments also apply to the fermionic contributions to the gauge current (where the derivative w.r.t.\ $y$ has been replaced by a division by $y$). In summary, the second term in (\ref{eq:bdos}) can be rewritten as (for further details see  \cite{Moitra:2018jqs})
\begin{equation}\label{eq:2donguage}
S_\sigma = -\pi^2 v_2 \int dx dy \sqrt{g} \left( g^{yy}\right)^2 j_y ^A \partial_x^{-2} j^A _y
\end{equation}

Now, we would like to compare this to the results of section \ref{sec:gravitini}. To this end, we replace the presently used 2d metric by one with the $AdS_2$ radius restored and an additional overall sign ($ z= x+ \text{i} y$)
\begin{equation}
ds^2 = -v_1^2 \frac{dzd\bar{z}}{y^2} .
\end{equation}
This changes, at most, a numerical factor in front of  (\ref{eq:2doneps}), (\ref{eq:2doneta}) and (\ref{eq:2donguage}). To make contact with the 4d near horizon $AdS_2$ factor in  (\ref{eq:4bh}) with $U(r) = v_1/r$ (and $b=v_2$, $b^\prime =0$) we perform the following coordinate transformation (including a Wick rotation)
\begin{equation}\label{eq:wicketal}
x = -\frac{\text{i}}{v_1} t \,\,\, ,\,\,\, y = \frac{v_1}{r} .
\end{equation}
Then energy momentum conservation (\ref{eq:emconJT}) matches the one obtained in the near horizon dimensional reduction (\ref{eq:emconDR}). This motivates us to associate the involved energy momentum tensors (up to an overall factor which would not change conservation laws). Indeed, applying (\ref{eq:wicketal}) on the corresponding part of the on-shell action $S_\epsilon$ in  (\ref{eq:2doneps}) matches the 4d near horizon result (\ref{eq:gravos}). The same observation holds for the $U(1)$ currents with conservation laws (\ref{eq:abelconJT}) respectively (\ref{eq:abelconNH}). The onshell actions (\ref{eq:2donguage}) and (\ref{eq:abelon}) agree as well.

For the gravitini sector the situation is more complicated. Performing a Wick rotation on spinorcomponents (such as the supercurrent) can be more involved
(see e.g.\ \cite{Osterwalder:1973kn,Nicolai:1978vc,vanNieuwenhuizen:1996tv}). In our setup, where we have projected everything to one component spinors the problem shows up as follows. Performing the transformation (\ref{eq:wicketal}) on the equations (\ref{eq:2dscurcon}) as well as on their complex conjugates will result in four equations which are not anymore pairwise related by complex conjugation. We proceed as follows. We just perform (\ref{eq:wicketal}) on the two equations written explicitly in (\ref{eq:2dscurcon}), resulting in
\begin{align}
\partial_t S^t_{\bar{+}}+\partial_r S^r_{\bar{+}} -\frac{r}{2v_1^2}\left( S_{\bar{+}}^t -S_{-}^t\right) -\frac{1}{2r} S_-^r &=0 ,\label{eq:wick1}
\\
\partial_t S^t_{\bar{-}}+\partial_r S^r_{\bar{-}} +\frac{r}{2v_1^2}\left( S_{\bar{-}}^t -S_{+}^t\right) -\frac{1}{2r} S_+^r &= 0 . \label{eq:wick2}
\end{align}
These, we compare to (\ref{4dsupercurrentconservation1}) and its complex conjugate.
This suggests the following association 
\begin{align}
&S_{\bar{+}}^t = \text{i} J_1 ^t\,\,\, ,\,\,\, S_{\bar{+}}^r = \text{i} J_1 ^r\,\,\, ,
\,\,\, S_+ ^t = \text{i} J_1 ^{t*}\,\,\, ,\,\,\, S_+ ^r = \text{i} J_1 ^{r*},\nonumber\\ &S_{\bar{-}}^t = - J_2 ^t\,\,\, ,\,\,\, S_{\bar{-}}^r =  J_2 ^r\,\,\, ,\,\,\,
S_- ^t = -J_2^{t*}\,\,\, ,\,\,\, S_{-}^r = J_2^{r*} .\label{eq:sucurcor}
\end{align}
Then (\ref{4dsupercurrentconservation2}) and its complex conjugate should map the set of conservation laws obtained from complex conjugation of (\ref{eq:wick1}) and (\ref{eq:wick2}). This indeed hapens if we apply the following rules for taking the complex conjugate of (\ref{eq:wick1}) and (\ref{eq:wick2})  ($\alpha \in \left\{ t,r\right\}$),
\begin{equation}\label{eq:strangecc}
\left(S_+ ^\alpha\right)^* = S_{\bar{-}}^\alpha \,\,\, ,\,\,\, \left( S_- ^\alpha\right)^* = S_{\bar{+}}^\alpha\,\,\, ,\,\,\, \partial_t ^* = -\partial_t .
\end{equation}
The relation between the current components is the same as it would be without performing a Wick rotation. Therefore one should take the complex conjugate of the coordinate transformation (\ref{eq:wicketal}) justifying the last assignment in 
(\ref{eq:strangecc}). Note also, that the prescription (\ref{eq:strangecc}) does not apply to the right hand sides of (\ref{eq:sucurcor}). That means in particular that after the replacement (\ref{eq:sucurcor}) the onshell action (\ref{eq:2doneps}) is not manifestly real anymore. Therefore we add its complex conjugate by hand. Finally, we arrive at
\begin{multline}\label{eq:wickon}
S_{\eta\bar{\eta}} \sim \int du dr \left[ \left( J_1^r -\text{i} J_2 ^{r*}\right)^* \partial_t^{-1} \left( J_1^t + \text{i} J_2 ^{t*}\right)\right. + \\ +\left. \left( J_1^r +\text{i} J_2 ^{r*}\right)^* \partial_t^{-1} \left( J_1^t - \text{i} J_2 ^{t*}\right) +\text{c.c.}\right] ,
\end{multline}
where now complex conjugation relates 4d components ($\left(J_A^\mu\right)^* = J_A^{\mu*}$). To compare with results from section \ref{sec:gravitini} we impose (\ref{eq:conrest}) which removes the first contribution to (\ref{eq:wickon}).
Further we notice that (\ref{4dsupercurrentconservation1}) and (\ref{4dsupercurrentconservation2}) imply
\begin{equation}
\left(\frac{2}{r}\partial_r +\frac{1}{r^2}\right) \left(J_1^r + \text{i} J_2 ^{r*}\right) = -\frac{2}{r}\partial_t \left( J_1^t + \text{i}J_2^{t*}\right)
\end{equation}
giving rise to a contact term in the near horizon limit. Hence, our expressions 
(\ref{eq:gravitiniout}) and (\ref{eq:wickon}) match within the given restriction (\ref{eq:conrest}).

\section{Discussion\label{ref:dis}}

\textbf{Summary}\\
In the first half of the paper we embedded the solution of \cite{Klemm,Hristov:2010ri} into a supergravity solution with the same amount of susy encompassing a hypermultiplet. This requires the choice of a moment map (and a corresponding Killing vector on the quaternion manifold) and choice of vacuum expectation values for the four hyperscalars. As a next step the dimensional reduction (in s-wave approximation) of the supergravity theory  is performed in the near horizon limit, hence on $AdS_2\times S^2$. To be more exact, we include fluctuations of photon, metric, gravitini and of the matter multiplet. For the latter we only let half of the hypermultiplet fluctuate, namely $u,v$ and one projection of the hyperinos, such that we acquire a proper two-dimensional multiplet. Contributions containing the background magnetic fieldstrength or the angular components of the spin connection drop out
due to the spherical integration, the BPS conditions and the choice of projection, such that a fully two-dimensional theory is furnished. The effective two-dimensional cosmological constant is given by a linear combination of the magnetic charge and the FI constants. We observe the dilaton coupled to metric fluctuations, an electric field strength term, an electric field strength to dilaton coupling and gravitini fluctuations coupled to the dilatini. The spherical fluctuations $h_{\theta \theta}, \psi_{\theta 1},\psi_{\theta 2}$ constitute the dilaton multiplet. Furthermore, we also see that deviations from pure JT supergravity occur due to additional source to dilaton multiplet couplings. We also calculate the four-point function in a dual CFT following the general prescription \cite{Liu:1998ty} within the same limits as discussed in \cite{Nayak:2018qej}. \\
In the second half of the paper we repeat the construction of  \cite{Forste:2017apw} for $\mathcal{N}=(2,2)$ Euclidean JT supergravity, while allowing the gravitini and graviphoton field strength to fluctuate. We focus on real supercurvature constraints, which take out half the degrees of freedom of the dilaton multiplet. Gauged matter can be added in form of a covariantly twisted chiral and anti-chiral multiplet and additional reality constraints due to the Euclidean signature. Somewhat unusually only a D-term is necessary to enable  gauged, massive matter fields. The masses are determined by the curvature, such that they agree with the dimensionally reduced  near-horizon theory. The D-term furthermore gives a linearized supergravity theory. Taking variational derivatives with respect to graviphoton, metric and gravitini furnishes symmetry currents. We show how the on-shell action of the matter coupled to the gravity multiplet fluctuations may be described via the boundary superspace two-point function. Then in combination with the super-Schwarzian up to quadratic order, we may integrate out the gravity multiplet fluctuations, such that we end up with a four point function described in terms of the currents. 
We compare our results with the computation obtained in the near horizon calculation of the four dimensional theory in which next to leading corrections to the $S^2$ radius have been taken into account. When expressed in terms of energy-momentum tensor and gauge current the results match. For the contribution containing the supercurrent the situation is a bit more involved. The limit in which only corrections to the $S^2$ radius are considered is not compatible with BPS conditions. Supercurrents are only conserved if we impose an additional projection. Up to terms vanishing under that projetion results from integrating out the fermionic super-Schwarzian mode match the four dimensional calculation. 
\\
\noindent \textbf{$\mathbf{\mathcal{N}=(2,2)}$ JT Quantum Supergravity}\\
As mentioned above, $\mathcal{N}=(2,2)$ JT supergravity encompasses a larger space of options then might be guessed when just performing the s-wave reduction. The dimensionally reduced theory naturally assumes a real dilaton and hence half the degrees of freedom available to the most general two-dimensional theory. A priori the two-dimensional theory might use $\Phi(R+\alpha)$, with $\alpha$ being a general complex number. This would fix the graviphoton to a specific background value. It would be interesting to explore the full range of solutions of this theory.\\
This is especially interesting with respect to the results of \cite{Saad:2019lba}, \cite{Stanford:2019vob}. By use of \cite{Mirzakhani:2006fta}, \cite{Saad:2019lba} calculated exact partition functions for JT gravity with arbitrary genus and arbitrary number of asymptotic boundaries. The partition functions are (non-uniquely) non-perturbatively completed by a genus expansion of a specific matrix integral. While \cite{Stanford:2019vob} extended these results to the $\mathcal{N}=1$ case\footnote{Supersymmetric extensions of JT gravity have been considered in this context also in \cite{Johnson:2020heh,Johnson:2020exp,Mertens:2020pfe}. }, it would be interesting to see the extension to $\mathcal{N}=2$ for the aforementioned reasons, although it is not quite clear how feasible this is.\\
\noindent \textbf{Localization}\\
As we have constructed a Euclidean off-shell formulation of an $\mathcal{N}=2$ supergravity theory coupled to matter, it is natural to consider localization techniques. For the general Schwarzian theory this was performed in \cite{Stanford:2017thb} for the bosonic case and also $\mathcal{N}=1$ and $\mathcal{N}=2$. It would be interesting to first of all, localize the minimal sugra theory on the $AdS_2$ background. The assumption would be that this should in leading order match the Schwarzian result for the partition function with differences perhaps arising in higher order corrections. Then one might attempt to perform this while also including the chiral twisted multiplet.\\
\noindent \textbf{Other Settings}\\
We have chosen the specific background of \cite{Klemm,Hristov:2010ri} as the near horizon enhancement matches with the two-dimensional theory first presented in \cite{Forste:2017apw} and for the fact that the $AdS_4$ asymptotics of \cite{Klemm,Hristov:2010ri} allow for the construction of a four-dimensional four point function via the $AdS/CFT$ dictionary. However, the most common four dimensional BPS solutions exhibit Minkowski asymptotics with near horizon enhancement to full BPS. It would be interesting to try and understand if these kind of solutions can be described via super-Schwarzian asymptotics.\\
\section*{Acknowledgements}  
We would like to thank Max Wiesner for collaboration on this project at the initial stage. Preliminary results had been reported in \cite{wiesner}. We would also like to thank Hans Jockers and Jun Nian for numerous discussions and Kyril Hristov for correspondence.
This work was supported by ``Bonn-Cologne
Graduate School for Physics and Astronomy'' (BCGS).

\begin{appendix}
\section{Four-Dimensional Supergravity Conventions\label{ap:conv}}
 
 In this appendix, we summarize the most important conventions taken from \cite{Andrianopoli:1996cm,Hristov:2012bk}.
 We use the mostly minus form of the Minkowski metric $\eta_{ab}=\text{diag}(1,-1,-1,-1)$.
 The flat space Dirac algebra of the $\gamma$-matrices is
\begin{align}
    \left\{\gamma_a, \gamma_b\right\}\equiv2\eta_{ab}\,,
\end{align}
with $\gamma_{ab}$ given by the commutator 
\begin{equation}
    \gamma_{ab}\equiv\frac{1}{2}\left[\gamma_a,\gamma_b\right]\,.  
\end{equation}
The chirality matrix is defined as
\begin{equation}\label{gamma5}
    \gamma_5\equiv -\text{i}\gamma_0\gamma_1\gamma_2\gamma_3=\text{i}\gamma^0\gamma^1\gamma^2\gamma^3\,.
\end{equation}
The $\gamma$ matrices are chosen to be purely imaginary $\left(\gamma_\mu\right)^*=-\gamma_\mu$ and furthermore
\begin{equation}
    \gamma_0^\dagger=\gamma_0\,,\;\;\; \gamma_0\gamma_i^\dagger \gamma_0=\gamma_i\,,\;\;\; \gamma_5^\dagger=\gamma_5 \,,\;\;\; i=1,2,3\,.
\end{equation}
 The $\gamma_5$ eigenvalues of the fermions  are
\begin{align}
    \gamma_5\left(\begin{matrix}\psi^\mu_A \\ \lambda^{iA} \\ \zeta_\alpha \end{matrix}\right)&=\left(\begin{matrix}\psi^\mu_A \\ \lambda^{iA} \\ \zeta_\alpha \end{matrix}\right) \,,\\
    \gamma_5\left(\begin{matrix}\psi^{\mu A} \\ \lambda^{i}_{A} \\ \zeta^\alpha \end{matrix}\right)&=-\left(\begin{matrix}\psi^{\mu A} \\ \lambda^{i}_{A} \\ \zeta^\alpha \end{matrix}\right) \,,
\end{align}
where $\psi^\mu_A$ is the gravitino, $\lambda^{iA}$ the gaugino and $\zeta_\alpha$ the hyperino. For this choice of $\gamma_5$
for chiral fermions we get
\begin{equation}
    \lambda_A^{\ast}=\lambda^{A}\,,\,\,\, \psi_{\mu A}^\ast = \psi_\mu ^A\,,\,\,\, \zeta_\alpha^\ast = \zeta^\alpha.
\end{equation}\\
In terms of the Pauli matrices the representation of the $\gamma$-matrices is
\begin{equation}
    \gamma^0=\left(\begin{matrix}0 &\sigma^2\\\sigma^2 & 0 \end{matrix}\right)\,,\;\gamma^1=\left(\begin{matrix}\text{i}\sigma^3 &0\\ 0&\text{i}\sigma^3 \end{matrix}\right)\,,\;\ \gamma^2=\left(\begin{matrix}0 &-\sigma^2\\\sigma^2 & 0 \end{matrix}\right)\,,\; \gamma^3=\left(\begin{matrix}-\text{i}\sigma^1&0\\0&-\text{i}
    \sigma^1 \end{matrix}\right)\,,
\end{equation}
where  $\sigma^i\,,\;i=1,2,3$ denote the Pauli matrices,
\begin{equation}
    \left(\sigma^1\right)_A{}^B=\left(\begin{matrix}0&1\\1&0\end{matrix}\right)\,,\;\;\; \left(\sigma^2\right)_A{}^B=\left(\begin{matrix}0&-\text{i}\\
    \text{i} &0\end{matrix}\right)\,,\;\;\; \left(\sigma^3\right)_A{}^B=\left(\begin{matrix}1&0\\0&-1\end{matrix}\right)\,.
\end{equation}
The $SU(2)$ indices $A,B$ are raised and lowered via the antisymmetric matrix 
$$\epsilon_{A B} =\left(\begin{matrix}0&1\\ -1&0\end{matrix}\right)\, ,\,\,\,  \epsilon^{AB}=\left(\begin{matrix}0&1\\-1&0\end{matrix}\right)$$ 
    such that we get 
\begin{equation}
    \left(\sigma^1\right)_{AB}=\left(\begin{matrix}1&0\\0&-1\end{matrix}\right)\,,\;\left(\sigma^2\right)_{AB}=\left(\begin{matrix}-\text{i}&0\\0&-\text{i}\end{matrix}\right)\,,\;\left(\sigma^3\right)_{AB}=\left(\begin{matrix}0&-1\\-1&0\end{matrix}\right)
\end{equation}
and
\begin{equation}
        \left(\sigma^1\right)^{AB}=\left(\begin{matrix}-1&0\\0&1\end{matrix}\right)\,,\;\;\;\left(\sigma^2\right)^{AB}=\left(\begin{matrix}-\text{i}&0\\0&-\text{i}\end{matrix}\right)\,,\;\;\;\left(\sigma^3\right)^{AB}=\left(\begin{matrix}0&1\\1&0\end{matrix}\right)\, .
\end{equation}
In the hypermultiplet sector the indices $\alpha,\beta$ are raised and lowered via the antisymmetric symplectic matrix $\mathbb{C}_{\alpha \beta}$
\begin{align}
    \mathbb{C}_{\alpha \beta}=\left(\begin{matrix}0&-1\\1&0\end{matrix}\right)\,.
\end{align}
With the  charge conjugation matrix 
\begin{equation}
    C=\text{i}\gamma_0\,,
\end{equation}
and 
\begin{equation}
    \overline{\lambda_A}=\text{i} \left(\lambda_A\right)^T\gamma_0\,,\;\,.
\end{equation}
chiral fermions satisfy
\begin{equation}
    \left(\overline{\lambda_A}\right)^*=\overline{\lambda^A}\,. 
\end{equation}
\end{appendix}


\begin{thebibliography}{99}
%
\bibitem{Nayak:2018qej}
  P.~Nayak, A.~Shukla, R.~M.~Soni, S.~P.~Trivedi and V.~Vishal,
  JHEP {\bf 1809} (2018) 048
  doi:10.1007/JHEP09(2018)048
  [arXiv:1802.09547 [hep-th]].
  %
  \bibitem{Moitra:2018jqs}
  U.~Moitra, S.~P.~Trivedi and V.~Vishal,
  JHEP {\bf 1907} (2019) 055
  doi:10.1007/JHEP07(2019)055
  [arXiv:1808.08239 [hep-th]].
%
\bibitem{Maldacena:1998uz}
  J.~M.~Maldacena, J.~Michelson and A.~Strominger,
  JHEP {\bf 9902} (1999) 011
  doi:10.1088/1126-6708/1999/02/011
  [hep-th/9812073].
  %
\bibitem{Almheiri:2014cka}
  A.~Almheiri and J.~Polchinski,
  JHEP {\bf 1511} (2015) 014
  doi:10.1007/JHEP11(2015)014
  [arXiv:1402.6334 [hep-th]].
  %
\bibitem{Jackiw:1984je}
  R.~Jackiw,
  Nucl.\ Phys.\ B {\bf 252} (1985) 343.
%
\bibitem{Teitelboim:1983ux}
  C.~Teitelboim,
  Phys.\ Lett.\  {\bf 126B} (1983) 41.
%
\bibitem{Jensen:2016pah}
  K.~Jensen,
  Phys.\ Rev.\ Lett.\  {\bf 117} (2016) no.11,  111601
  [arXiv:1605.06098 [hep-th]].
%
\bibitem{Maldacena:2016upp}
  J.~Maldacena, D.~Stanford and Z.~Yang,
  PTEP {\bf 2016} (2016) no.12,  12C104
  [arXiv:1606.01857 [hep-th]].
%
\bibitem{Engelsoy:2016xyb}
  J.~Engelsöy, T.~G.~Mertens and H.~Verlinde,
  JHEP {\bf 1607} (2016) 139
  doi:10.1007/JHEP07(2016)139
  [arXiv:1606.03438 [hep-th]].
%
\bibitem{Cvetic:2016eiv}
  M.~Cveti\v{c} and I.~Papadimitriou,
  JHEP {\bf 1612} (2016) 008
   Erratum: [JHEP {\bf 1701} (2017) 120]
  [arXiv:1608.07018 [hep-th]].
%
\bibitem{Sachdev:1992fk}
  S.~Sachdev and J.~Ye,
  Phys.\ Rev.\ Lett.\  {\bf 70} (1993) 3339
  [cond-mat/9212030].
%
\bibitem{Kitaev}
A. Kitaev, “A Simple Model Of Quantum Holography,” talks at KITP, April
7, 2015 and May 27, 2015,
http://online.kitp.ucsb.edu/online/entangled15/kitaev/, 
http://online.kitp.ucsb.edu/online/entangled15/kitaev2/.
%
\bibitem{Maldacena:2016hyu}
  J.~Maldacena and D.~Stanford,
  Phys.\ Rev.\ D {\bf 94} (2016) no.10,  106002
  [arXiv:1604.07818 [hep-th]].
%
\bibitem{Sarosi:2017ykf}
G.~Sárosi,
PoS \textbf{Modave2017} (2018), 001
doi:10.22323/1.323.0001
[arXiv:1711.08482 [hep-th]].
%
\bibitem{Rosenhaus:2018dtp}
V.~Rosenhaus,
J. Phys. A \textbf{52} (2019), 323001
doi:10.1088/1751-8121/ab2ce1
[arXiv:1807.03334 [hep-th]].
%
\bibitem{Chamseddine:1991fg}
  A.~H.~Chamseddine,
  Phys.\ Lett.\ B {\bf 258} (1991) 97.
  doi:10.1016/0370-2693(91)91215-H
%
\bibitem{Fu:2016vas}
  W.~Fu, D.~Gaiotto, J.~Maldacena and S.~Sachdev,
  Phys.\ Rev.\ D {\bf 95} (2017) no.2,  026009
   Addendum: [Phys.\ Rev.\ D {\bf 95} (2017) no.6,  069904]
  doi:10.1103/PhysRevD.95.069904, 10.1103/PhysRevD.95.026009
  [arXiv:1610.08917 [hep-th]].
%
\bibitem{Astorino:2002bj}
M.~Astorino, S.~Cacciatori, D.~Klemm and D.~Zanon,
Annals Phys. \textbf{304} (2003), 128-144
doi:10.1016/S0003-4916(03)00008-3
[arXiv:hep-th/0212096 [hep-th]].
\bibitem{Forste:2017kwy}
  S.~F{\"o}rste and I.~Golla,
  Phys.\ Lett.\ B {\bf 771} (2017) 157
  doi:10.1016/j.physletb.2017.05.039
  [arXiv:1703.10969 [hep-th]].
%
\bibitem{Forste:2017apw}
  S.~F{\"o}rste, J.~Kames-King and M.~Wiesner,
  JHEP {\bf 1803} (2018) 028
  doi:10.1007/JHEP03(2018)028
  [arXiv:1712.07398 [hep-th]].
%
\bibitem{Stanford:2017thb}
D.~Stanford and E.~Witten,
JHEP \textbf{10} (2017), 008
doi:10.1007/JHEP10(2017)008
[arXiv:1703.04612 [hep-th]].
%
\bibitem{Mertens:2017mtv}
T.~G.~Mertens, G.~J.~Turiaci and H.~L.~Verlinde,
JHEP \textbf{08} (2017), 136
doi:10.1007/JHEP08(2017)136
[arXiv:1705.08408 [hep-th]].
%
\bibitem{Kanazawa:2017dpd}
T.~Kanazawa and T.~Wettig,
JHEP \textbf{09} (2017), 050
doi:10.1007/JHEP09(2017)050
[arXiv:1706.03044 [hep-th]].
%
\bibitem{Murugan:2017eto}
J.~Murugan, D.~Stanford and E.~Witten,
JHEP \textbf{08} (2017), 146
doi:10.1007/JHEP08(2017)146
[arXiv:1706.05362 [hep-th]].
%
\bibitem{Yoon:2017gut}
J.~Yoon,
JHEP \textbf{10} (2017), 172
doi:10.1007/JHEP10(2017)172
[arXiv:1706.05914 [hep-th]].
%
\bibitem{neues}
N.~Sannomiya, H.~Katsura and Y.~Nakayama,
Phys. Rev. D \textbf{95} (2017) no.6, 065001
doi:10.1103/PhysRevD.95.065001
[arXiv:1612.02285 [cond-mat.str-el]].


\bibitem{Peng:2017spg}
C.~Peng, M.~Spradlin and A.~Volovich,
JHEP \textbf{10} (2017), 202
doi:10.1007/JHEP10(2017)202
[arXiv:1706.06078 [hep-th]].
%
\bibitem{Bulycheva:2018qcp}
K.~Bulycheva,
JHEP \textbf{04} (2018), 036
doi:10.1007/JHEP04(2018)036
[arXiv:1801.09006 [hep-th]].
 %
%
\bibitem{Peng:2018zap}
C.~Peng,
JHEP \textbf{12} (2018), 065
doi:10.1007/JHEP12(2018)065
[arXiv:1805.09325 [hep-th]]
%
\bibitem{Chang:2018sve}
C.~M.~Chang, S.~Colin-Ellerin and M.~Rangamani,
JHEP \textbf{10} (2018), 157
doi:10.1007/JHEP10(2018)157
[arXiv:1806.09903 [hep-th]].
%
\bibitem{Cardenas:2018krd}
M.~Cárdenas, O.~Fuentealba, H.~A.~González, D.~Grumiller, C.~Valcárcel and D.~Vassilevich,
JHEP \textbf{11} (2018), 077
doi:10.1007/JHEP11(2018)077
[arXiv:1809.07208 [hep-th]].
%
\bibitem{Berkooz:2020xne}
M.~Berkooz, N.~Brukner, V.~Narovlansky and A.~Raz,
[arXiv:2003.04405 [hep-th]].
%
\bibitem{Peng:2020euz}
C.~Peng and S.~Stanojevic,
[arXiv:2006.13961 [hep-th]].
%
\bibitem{Castro:2018ffi}
  A.~Castro, F.~Larsen and I.~Papadimitriou,
  JHEP {\bf 1810} (2018) 042
  doi:10.1007/JHEP10(2018)042
  [arXiv:1807.06988 [hep-th]].
%
\bibitem{Moitra:2019bub}
  U.~Moitra, S.~K.~Sake, S.~P.~Trivedi and V.~Vishal,
  arXiv:1905.10378 [hep-th].
%
\bibitem{Liu:1998ty}
  H.~Liu and A.~A.~Tseytlin,
  Phys.\ Rev.\ D {\bf 59} (1999) 086002
  doi:10.1103/PhysRevD.59.086002
  [hep-th/9807097].
%
\bibitem{Klemm}
S.~L.~Cacciatori and D.~Klemm,
JHEP \textbf{01} (2010), 085
doi:10.1007/JHEP01(2010)085
[arXiv:0911.4926 [hep-th]].
\bibitem{Hristov:2010ri}
  K.~Hristov and S.~Vandoren,
  JHEP {\bf 1104} (2011) 047
  doi:10.1007/JHEP04(2011)047
  [arXiv:1012.4314 [hep-th]].
%
\bibitem{Hristov:2010eu}
  K.~Hristov, H.~Looyestijn and S.~Vandoren,
  JHEP {\bf 1008} (2010) 103
  doi:10.1007/JHEP08(2010)103
  [arXiv:1005.3650 [hep-th]].
%
\bibitem{Romans:1991nq}
  L.~J.~Romans,
  Nucl.\ Phys.\ B {\bf 383} (1992) 395
  doi:10.1016/0550-3213(92)90684-4
  [hep-th/9203018].
%
\bibitem{Andrianopoli:1996cm}
  L.~Andrianopoli, M.~Bertolini, A.~Ceresole, R.~D'Auria, S.~Ferrara, P.~Fre and T.~Magri,
  J.\ Geom.\ Phys.\  {\bf 23} (1997) 111
  doi:10.1016/S0393-0440(97)00002-8
  [hep-th/9605032].
%
\bibitem{Freedman:2012zz}
  D.~Z.~Freedman and A.~Van Proeyen,
  {\it Supergravity}, Cambridge, UK: Cambridge Univ. Press, 2015, {\scriptsize ISBN:} 9781139368063, 9780521194013
%
\bibitem{Ortin:2015hya}
  T.~Ortin,
  {\it Gravity and Strings},
  Cambridge Monographs on Mathematical Physics,
  Cambridge University Press, 2015, {\scriptsize ISBN:} 9780521768139, 9780521768139, 9781316235799,
  doi:10.1017/CBO9781139019750
%
\bibitem{Hristov:2012bk}
  K.~Hristov,
  arXiv:1207.3830 [hep-th].
%
\bibitem{Productspace}
B.~Gouteraux, J.~Smolic, M.~Smolic, K.~Skenderis and M.~Taylor,
JHEP \textbf{01} (2012), 089
doi:10.1007/JHEP01(2012)089
[arXiv:1110.2320 [hep-th]].
\bibitem{n=2sugra}
L.~Andrianopoli, M.~Bertolini, A.~Ceresole, R.~D'Auria, S.~Ferrara, P.~Fre and T.~Magri,
J. Geom. Phys. \textbf{23} (1997), 111-189
doi:10.1016/S0393-0440(97)00002-8
[arXiv:hep-th/9605032 [hep-th]].
%
\bibitem{Witten:1998qj}
E.~Witten,
Adv. Theor. Math. Phys. \textbf{2} (1998), 253-291
doi:10.4310/ATMP.1998.v2.n2.a2
[arXiv:hep-th/9802150 [hep-th]].
%
\bibitem{Skenderis:2002wp}
K.~Skenderis,
Class. Quant. Grav. \textbf{19} (2002), 5849-5876
doi:10.1088/0264-9381/19/22/306
[arXiv:hep-th/0209067 [hep-th]].
%
\bibitem{Hull:2008de}
C.~M.~Hull, U.~Lindstrom, L.~Melo dos Santos, R.~von Unge and M.~Zabzine,
JHEP \textbf{06} (2008), 031
doi:10.1088/1126-6708/2008/06/031
[arXiv:0805.3321 [hep-th]].
%
\bibitem{deWit:2011gk}
  B.~de Wit and M.~van Zalk,
  JHEP {\bf 1110} (2011) 050
  doi:10.1007/JHEP10(2011)050
  [arXiv:1107.3305 [hep-th]].
  %
\bibitem{Gates:1995du}
  S.~J.~Gates, Jr., M.~T.~Grisaru and M.~E.~Wehlau,
  Nucl.\ Phys.\ B {\bf 460} (1996) 579
  doi:10.1016/0550-3213(95)00648-6
  [hep-th/9509021].
 %
\bibitem{Freedman:1998tz}
D.~Z.~Freedman, S.~D.~Mathur, A.~Matusis and L.~Rastelli,
Nucl. Phys. B \textbf{546} (1999), 96-118
doi:10.1016/S0550-3213(99)00053-X
[arXiv:hep-th/9804058 [hep-th]].
 %
\bibitem{Osterwalder:1973kn}
K.~Osterwalder and R.~Schrader,
Phys. Rev. Lett. \textbf{29} (1972), 1423-1425
doi:10.1103/PhysRevLett.29.1423
%
\bibitem{Nicolai:1978vc}
H.~Nicolai,
Nucl. Phys. B \textbf{140} (1978), 294-300
doi:10.1016/0550-3213(78)90537-0
%
\bibitem{vanNieuwenhuizen:1996tv}
P.~van Nieuwenhuizen and A.~Waldron,
Phys. Lett. B \textbf{389} (1996), 29-36
doi:10.1016/S0370-2693(96)01251-8
[arXiv:hep-th/9608174 [hep-th]]
%
\bibitem{Saad:2019lba}
P.~Saad, S.~H.~Shenker and D.~Stanford,
[arXiv:1903.11115 [hep-th]].
%
\bibitem{Stanford:2019vob}
D.~Stanford and E.~Witten,
[arXiv:1907.03363 [hep-th]].
5
\bibitem{Mirzakhani:2006fta}
M.~Mirzakhani,
Invent. Math. \textbf{167} (2006) no.1, 179-222
doi:10.1007/s00222-006-0013-2
%
\bibitem{Johnson:2020heh}
C.~V.~Johnson,
[arXiv:2005.01893 [hep-th]]
%
\bibitem{Johnson:2020exp}
C.~V.~Johnson,
[arXiv:2006.10959 [hep-th]].
%
\bibitem{Mertens:2020pfe}
T.~G.~Mertens,
[arXiv:2007.00998 [hep-th]].
%
\bibitem{wiesner}
M.\ Wiesner, {\it ``Dilaton Supergravity and Near-Horizon Dynamicss of Supersymmetric Black Holes''}, Master Thesis, Bonn University, September 2018.
  \end{thebibliography}
\end{document}